\documentclass[twocolumn,aps,prb]{revtex4}
\usepackage{amsmath,empheq}
\usepackage{amssymb}
\usepackage{mathrsfs}  
\usepackage{amsfonts}
\usepackage{booktabs}
\usepackage{physics}
\usepackage{graphicx} 
\usepackage{subfigure} 
\usepackage{epsfig}
\usepackage{color}
\usepackage{epstopdf}
 \usepackage{rotating}
 \usepackage{appendix}
 \usepackage[colorlinks=true,linkcolor=blue,citecolor=blue,filecolor=blue,urlcolor=blue]{hyperref}

\begin{document}

 \title{Majorana chiral spin liquid in Mott insulating cuprates}
 \author{Jaime Merino}
\affiliation{Departamento de F\'isica Te\'orica de la Materia Condensada, Condensed Matter Physics Center (IFIMAC) and
Instituto Nicol\'as Cabrera, Universidad Aut\'onoma de Madrid, Madrid 28049, Spain}
\author{Arnaud Ralko}
\affiliation{Institut N\'eel, Universit\'e Grenoble Alpes et CNRS UPR2940, Grenoble 38042, France} 

\begin{abstract}
The large thermal Hall conductivity recently detected in Mott insulating cuprates has been attributed to 
 chiral neutral spin excitations. A quantum spin liquid with Majorana excitations,
 Chern number $\pm 4$ and large thermal Hall conductivity is found to be an excited state of a frustrated Heisenberg model on 
 the square lattice. 
 Using a Majorana mean-field theory and exact diagonalizations, 
 we explore two possible routes to achieve this chiral quantum spin liquid, an orbital effect of an applied magnetic field and spin orbit couplings as present in cuprates. In particular, we show how only the orbital magnetic field allows this topological phase to be the ground state, while it remains an excited state of the Majorana mean field under the Dzyaloshinskii-Moriya terms.
 We interpret the large thermal Hall effect 
 observed in Mott cuprates from their close proximity to a transition to a Majorana chiral quantum spin liquid which 
 can be induced by an external magnetic field.
 \end{abstract}
 \date{\today}

 \maketitle
 
{\em Introduction.} The pseudogap phase of the cuprates continues to provide unexpected behavior. Thermal Hall experiments 
have recently found a surprisingly large thermal Hall conductivity\cite{Taillefer2019}.
As doping is reduced below the critical doping $\delta< \delta^*$, a negative 
thermal Hall conductivity signal is observed, which becomes the largest 
when reaching the Mott insulator $\delta \approx 0$. The large absolute values 
and $T$-dependence of the thermal Hall conductivity in the undoped cuprate, 
La$_2$CuO$_4$, is very similar to observations in several spin liquid frustrated 
materials such as volborthite\cite{Watanabe2016} or $\alpha$-RuCl$_3$\cite{Kasahara2018a}. 
The half-quantization thermal Hall effect \cite{Kasahara2018b} observed in $\alpha$-RuCl$_3$
is interpreted in terms of Majorana edge modes arising in the 
Kitaev spin liquid under an external magnetic field. It is then an open
question whether the Mott and pseudogap phases of cuprates host unconventional 
neutral chiral excitations which can lead to the observed large thermal Hall effect.

Phonons have been identified as the main heat carriers in the thermal Hall	
effect observed in the pseudogap \cite{Taillefer2020a} and Mott insulating phases\cite{Taillefer2020b} 
of cuprates. The large values of the thermal Hall conductivity
indicates that phonons have non-zero chirality whose origin is yet 
to be explained. Since magnetic impurity effects and magnons have been discarded,
a possible scenario is that the paramagnetic phase of cuprates 
is a quantum spin liquid whose chirality is imprinted on the phonons 
through the spin-phonon coupling.  An intriguing theoretical 
possibility would be a chiral quantum spin liquid on the square lattice 
with Majorana fermions as elementary excitations, as in 
the Kitaev quantum spin liquid in the honeycomb lattice, relevant to $\alpha$-RuCl$_3$. 
Other related but different chiral quantum spin liquids with either bosonic\cite{Sachdev2019a,Sachdev2019b} or 
fermionic spinons\cite{Lee2019} have been proposed in undoped cuprates. The 
$d$-density wave \cite{HaiLee2019} has non-fractional electron constituents for $\delta \neq 0$ though.

Here, we show how a chiral spin liquid state with Majorana excitations is a good candidate for explaining the thermal Hall effect in Mott insulating 
cuprates. Using a Majorana representation of the $J_1$-$J_2$ Heisenberg model on the square lattice,
we find that a chiral spin liquid state breaking time-reversal symmetry 
with Majorana excitations emerges spontaneously. Such state which we denote as Majorana $\pi$-QSL has 
a large associated Chern number $\nu= \pm4$, leading to absolute values of the thermal Hall conductivity 
of the order $\sim k_B^2/\hbar$ as experimentally observed. However, 
this chiral spin liquid state is only an excited state of the system: the 
well known N\'eel, collinear and/or non-chiral disordered states being favored for different 
ranges of magnetic frustration.
Extending the Heisenberg model to account for the presence of an external magnetic field through the orbital effect and/or for a Dzyaloshinskii-Moriya (DM) spin orbit coupling, we explore whether 
the $\pi$-QSL becomes the absolute ground state of the system or not. 
Considering several compatible choices of DM vectors, we show how the 
$\pi$-QSL is lowered in energy by the DM but always 
remains an excited state of the system. 
Only the orbital magnetic field is able to turn the $\pi$-QSL into the absolute ground state, consistent with observations where the magnetic field is essential to externally trigger the transition. 
Hence, the combined effect of the DM and the magnetic field orbital
term can drive the relevant model for Mott insulating cuprates into 
the Majorana $\pi$-QSL phase which displays the large 
thermal Hall effect experimentally observed.

\begin{figure}
	\centering
	\includegraphics[width=0.35\textwidth,clip]{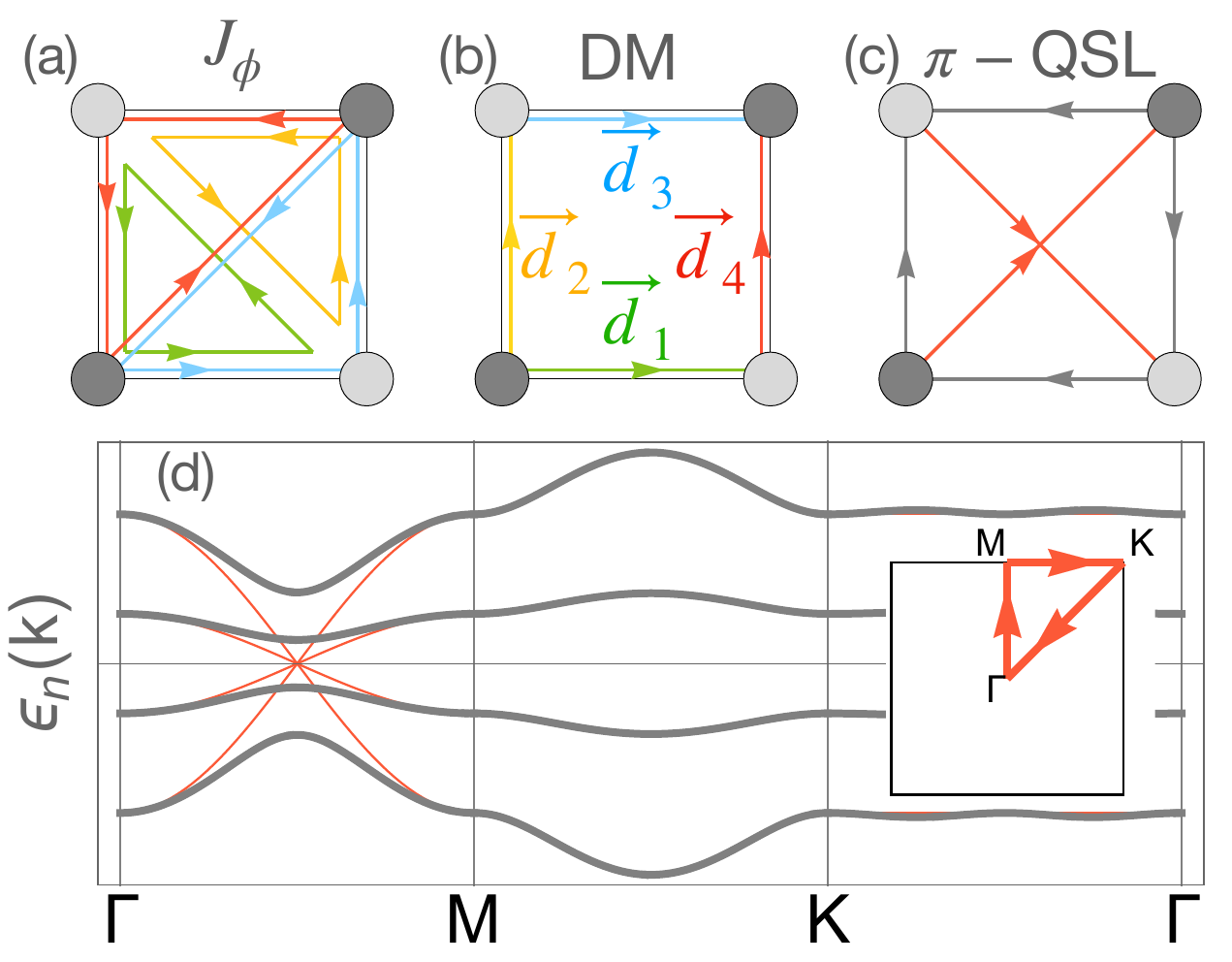}
	\caption{
	Majorana chiral quantum spin liquid state in the $J_1$-$J_2$ Heisenberg model on a square lattice. (a) and (b) Bond conventions used for the orbital magnetic field term and the Dzyaloshinskii-Moriya interactions. (c) The bond patterns of the MMFT lowest energy QSL, the $\pi$-QSL. The arrows indicate that the bond average is positive (negative) in the same (opposite) direction of a bond
	between two sites. (d) Excitation spectra of the 
	gapless $\pi$-QSL for $J_2=0$ (red) and of the gapped $\pi$-QSL 
	arising for $J_2 \neq 0$ ($J_2 =J_1/2$ in this plot). 
	The Brillouin zone with high symmetry points is displayed 
	as an inset.}
	\label{fig:disp}
\end{figure}

{\em Model and methods.} The simplest relevant model to describe the magnetic properties of undoped cuprates is the $J_1$-$J_2$  $S=1/2$ Heisenberg model on the square lattice:
 \begin{equation}
 H_\text{H}= J_1 \sum_{\langle i j \rangle}  {\bf S}_i \cdot {\bf S}_j  + J_2 \sum_{\langle \langle i j \rangle \rangle }   {\bf S}_i \cdot {\bf S}_j
 \end{equation}
where the first sum runs over nearest-neighbor sites and the second over next nearest neighbors. 

Electrons can couple to an external magnetic field $B$ through their orbital motion. A strong coupling expansion of the Hubbard model to $\mathcal{O}(t^3/U^2)$ leads to the additional chiral term\cite{Motrunich2006}:
\begin{equation}
H_\phi=J_\phi \sum_{\triangle} T_{ijk} = J_\phi \sum_{\triangle}   {\bf S}_i \cdot ({\bf S}_j  \times {\bf S}_k),
\end{equation}
where $\triangle$ denotes a triangle lying in the square plaquettes of area $A$ with vertices $ijk$ taken in an anticlockwise direction (see Fig.\ref{fig:disp}(b)). The three-spin-exchange coupling is $J_\phi=-{24 t_2 t_1^2 \over U^2} \sin(2 \pi \phi/\phi_0)$, with $\phi=B A/2$ and $\phi_0= h c / e$ the flux quantum. 

A second important ingredient is the spin-orbit coupling effect through the Dzyaloshinskii-Moriya (DM) interaction \cite{Dzyaloshinskii1958,Moriya1960a, 
Moriya1960b} defined as:
\begin{eqnarray}
    H_\text{DM} &=& \sum_{\langle i,j\rangle} {\bf D}_{ij} \cdot \left(  {\bf S}_i \cross {\bf S}_j \right)
\end{eqnarray}
where ${\bf D}_{ij}$ are DM vectors. The compatible vectors ${\bf D}_{ij}$ are generically defined by 4 unit vectors $\vec{d}_i$ with $i=1,2,3,4$ pointing in the different bond directions of a 4-site unit cell, as given in Fig.\ref{fig:disp}(b). We consider in this work several cases to explore the possibility of spin orbit coupling to lead to a chiral liquid state: SO compatible with cuprates like YBCO, LSCO-LTO, LSCO-LTT as defined in [\onlinecite{Sachdev2019a}], or SO corresponding to \textit{Rashba-like} and \textit{Dresselhaus-like} couplings [\onlinecite{Hotta2019}] 
(see Supplemental Material [\onlinecite{suppl}] for details).

We introduce the  Majorana representation of the spins consisting on four Majorana fermion operators $c, b^x, b^y, b^z$ per site 
as used by Kitaev \cite{Kitaev2006}  to exactly solve his spin model (see the Supplemental Material [\onlinecite{suppl}]).
In this representation, the spin operators are $S^\alpha_i= {i \over 2} b^\alpha_i c_i$,
where $\alpha=x,y, z$. A three-channel mean-field decoupling of the interaction terms $S^\alpha_i S^\beta_j$ in the Hamiltonian yields to 
\begin{eqnarray}
&+&\frac{1}{4} \left[ \langle b_i^\alpha  i c_i \rangle  b_j^\beta i c_j + \langle b_j^\beta i c_j \rangle b^\alpha_i i c_i -\langle b_i^\alpha i c_i \rangle \langle  b_j^\beta i c_j\rangle \right] \nonumber \\
&-&\frac{1}{4} \left[ \langle b_i^\alpha i b_j^\beta \rangle   c_i i c_ j + \langle c_i  i c_j \rangle b^\alpha_i i b^\beta_j - \langle b^\alpha_i i b^\beta_j \rangle \langle c_i i c_j \rangle \right]
\nonumber \\
&+&\frac{1}{4} \left[ \langle b_i^\alpha  i c_j \rangle  b_j^\beta i c_i + \langle b_j^\beta i c_i \rangle b^\alpha_i i c_j -\langle b_i^\alpha i c_j \rangle \langle  b_j^\beta i c_i\rangle \right] 
\end{eqnarray}
where the first three terms are associated with magnetic orders while the next three with spin liquid formation.

With such decoupling, our Majorana mean-field theory\cite{Ralko2020} (MMFT) allows to analyze on equal footing 
quantum spin liquids (QSL)($\langle {\bf S}_i \rangle =0$)  as well as magnetically ordered states. To do so, as done in [\onlinecite{Ralko2020}], we solve numerically the set of self consistent equations, together with a physical constraint connected to the number of particles per site that our theory has to fulfill.
\begin{figure}
	\centering
	\includegraphics[width=0.40\textwidth]{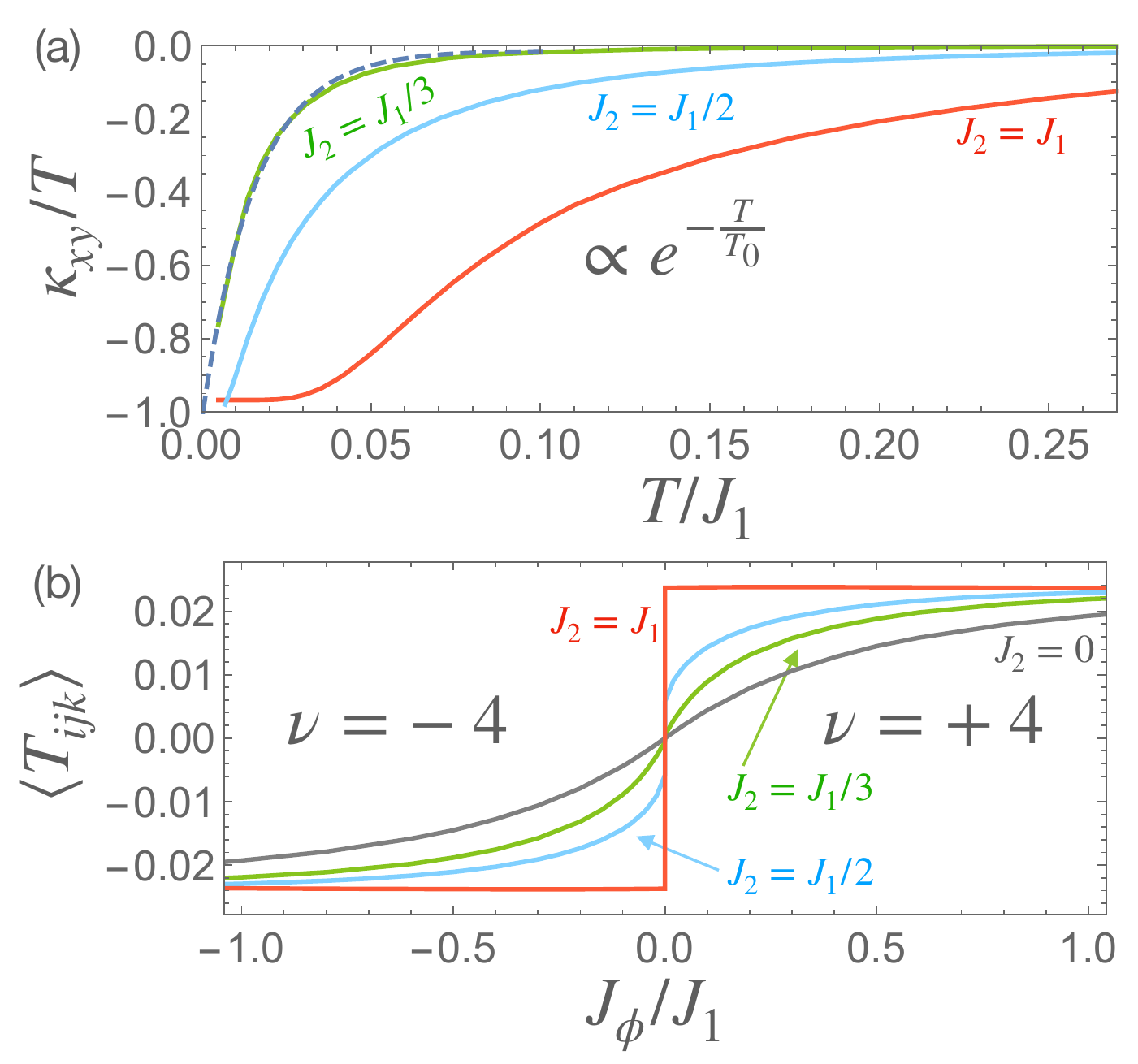}
	\caption{(a) Thermal Hall conductivity of the $\pi$-QSL on the square lattice. The $T$-dependence of $\kappa_{xy}/T$ in units of $k_B^2/\hbar$ is shown for different $J_2/J_1$. The $\pi$-QSL leads to large absolute values of $\kappa_{xy}/T \sim  -k_B^2 / \hbar$ as $T\rightarrow 0$ and an intermediate $T$-dependence $\kappa_{xy}/T \propto e^{-{T \over T_0}}$ consistent with observations in undoped cuprates. From such exponential fit to the $J_2 = J_1/3$ theoretical results we obtain a $T_0 \approx 18$ K (assuming $J_1 \sim 0.1$ eV), very close to experimental values, $T_0^{expt} \approx 16-18$ K for La$_2$CuO$_4$ and Sr$_2$CuO$_2$Cl$_2$. (b) Dependence of the chirality of the $\pi$-QSL state with $J_\phi$ obtained from MMFT. Dependence of the Chern number of the $\pi$-QSL state with $J_\phi$ for all $J_2$.} 
	\label{fig:the}
\end{figure}  

{\em Chiral QSL with Majorana excitations.} The  $J_1$-$J_2$ Heisenberg model on the square lattice sustains 
a chiral quantum spin liquid state with Majorana fermion excitations
(see Supplemental Material [\onlinecite{suppl}]). Among the possible QSL ansatze, the $\pi$-flux QSL ansatz or $\pi$-QSL from now on, shown Fig.\ref{fig:disp}(c) has the lowest energy in a finite range of $J_2$, consistent with 
Lieb's theorem \cite{Lieb1994}. The eight corresponding Majorana bands -- there are two sites per unit-cell\cite{sizeofunitcell}, labelled A and B -- are given by $\pm  3 \gamma({\bf k}), \pm \gamma({\bf k})$ (triply degenerate) with
\begin{equation*}
\gamma ( {\bf k}) = {J_1 \over 2} \sqrt{  c_1^2 ( \sin^2 k_x +\cos^2 k_y ) 
+ (2 {J_2 \over J_1} c_2 \cos k_x  \sin k_y)^2 }
\label{eq:gamma}
\end{equation*}
 where we have defined $c_1=\langle c_{i_A} i c_{j_B} \rangle= \langle b_{i_A}^\alpha i b_{j_B}^\alpha \rangle$ and $c_2=\langle c_{i_A} i c_{j_A} \rangle=\langle b_{i_A}^\alpha i b_{j_A}^\alpha \rangle $.
 For $J_2 = 0$, all these bands touch at the two Dirac points $(0,\pm \pi/2)$ in the Brillouin zone as shown in 
 Fig.\ref{fig:disp}(d), the $\pi$-QSL is gapless in this case. Moreover, it minimizes the total energy with $E=-0.3442 J_1$
much lower than the $0$-flux QSL ansatz with $E=-0.2462 J_1$.
The $\pm \pi$ Berry phases at the Dirac cones can lead to
non-trivial topology if the gap opens in a non-trivial manner. 
Indeed, this occurs when turning on a $J_2 \neq 0$, as shown in Fig.\ref{fig:disp}(d).
The gap opening lowers slightly the energy from  $E=-0.3442 J_1$ ($|c_1|=0.4790$) for $J_2=0$
to $E=-0.3443 J_1$  ($|c_1| = 0.4777$, $|c_2|=0.0516$) for $J_2=0.5$ and to $E=-0.3726 J_1$  ($|c_1| = 0.4189$, 
$|c_2|=0.2700$) for $J_2=J_1$. Moreover, a direct calculation of the topological invariant Chern number $\nu$ on the occupied Majorana bands\cite{Ralko2020} (with negative energy) shows that this state is a gapped topological $\pi$-QSL with $\nu = \pm 4$. 

In Fig. \ref{fig:the}(a) we show the thermal Hall conductivity of the 
$\pi$-QSL obtained from the expression:\cite{Qin2011}
\begin{equation}
{ \kappa_{xy} \over T} = {k_B^2 \over \hbar }  {1 \over 8 (k_B T)^3} \int d \epsilon { (\epsilon-\mu)^2 \over \cosh^2[\beta (\epsilon-\mu)/2]} \sigma_{xy}(\epsilon),
\label{eq:thcond}
\end{equation}
where $\sigma_{xy} (\epsilon) =  \sum_{n {\bf k}} \Omega^z_n({\bf k}) \theta( \epsilon-\epsilon_n({\bf k}))$. Note that the r.h.s. of Eq. \ref{eq:thcond} is half of the standard expression used for fermionic spinons since in our MMFT approach, thermal energy is transported by the Majoranas. In the limit $T \rightarrow 0$, we find  $\kappa_{xy}/T   \rightarrow -({k_B^2 \over \hbar})$ since each of the occupied bands carries a Chern number of $-1$. 
The overall absolute values and $T$-dependence of $\kappa_{xy}/T$ 
are consistent with observations (see Supplemental Material [\onlinecite{suppl}]). As temperature is raised and thermal excitations of the Majorana fermions occur, there is a cancellation between the Chern numbers of the bands above and below zero energy and $\kappa_{xy}/T \rightarrow 0$.

The gapped $\pi$-QSL found in the $J_1$-$J_2$ Heisenberg model 
for $J_2 \neq 0$ is doubly-degenerate since there are two possible senses 
of the bond amplitudes or chiralities, $\langle T_{ijk} \rangle \neq 0$, with the same energy.
Hence, the Chern number of the $\pi$-QSL can have two values, 
$\nu=\pm 4$, depending on the sign of the chirality as shown in Fig.\ref{fig:the}(b). It may seem that spontaneous symmetry breaking 
can occur since any infinitesimal $J_\phi \rightarrow 0^\pm$ 
selects either the $\nu=4$ or the $\nu=-4$ $\pi$-QSL solution. However,
our MMFT stability analysis below reveals that 
the $\pi$-QSL is only an excitation of the pure $J_1-J_2$ 
Heisenberg model. Under a finite $J_\phi$,
the $\pi$-QSL with the favorable sign of the chirality 
for the orientation of the applied $B$, becomes the ground state of the system.
This is consistent with experiments on the cuprates 
which find no thermal Hall effect when no 
magnetic field is applied implying no spontaneous 
time-reversal symmetry breaking ground state.

\begin{figure}[!h]
	\includegraphics[width=0.5\textwidth,clip]{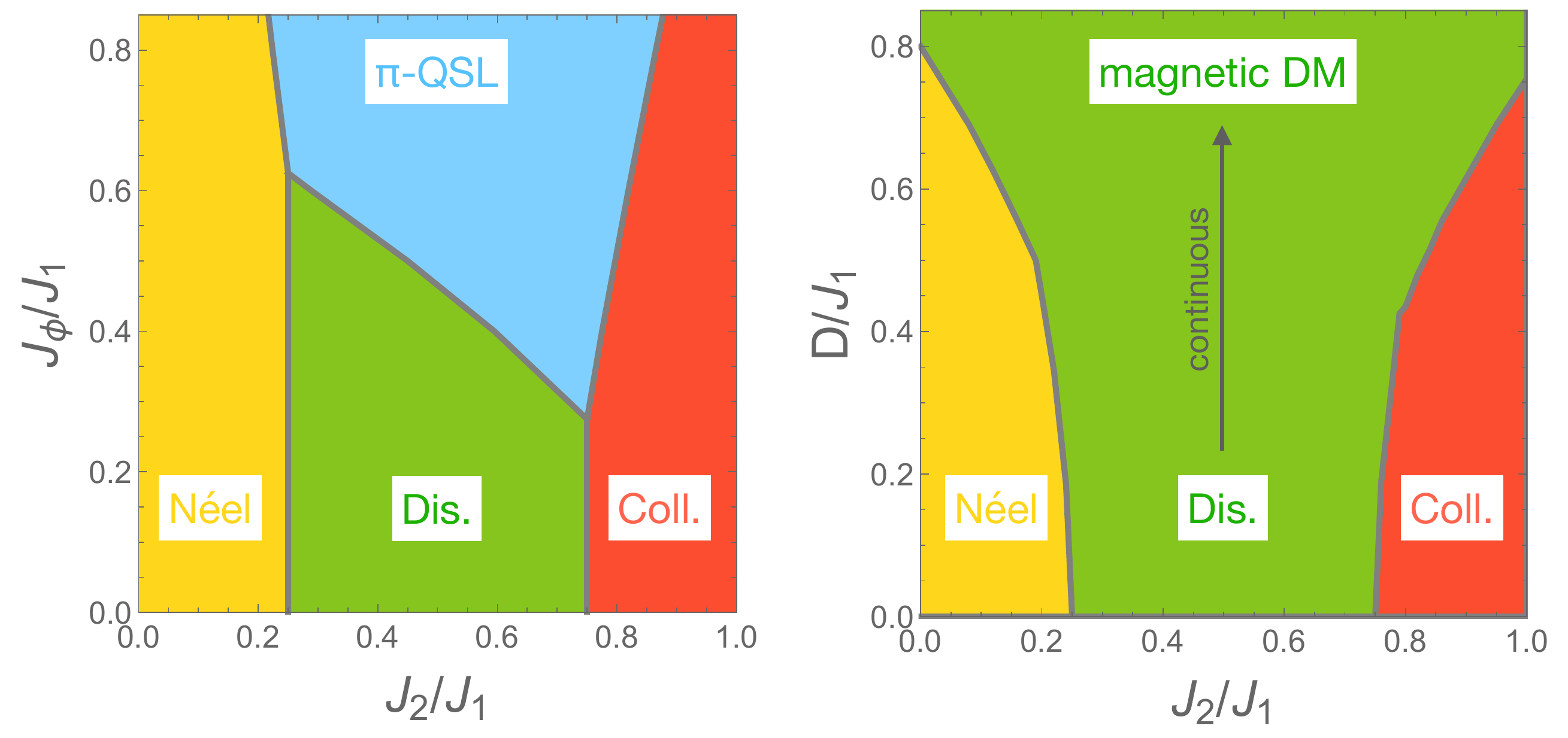}
    \caption{
    Phase diagrams of the  $J_\phi$ and $D$ models obtained from  Majorana mean field theory for cluster of $2 \times 32 \times 32$ sites. Left: orbital magnetic field, $J_\phi$, model. The intermediate phase (blue region) is the $\pi$-flux QSL with topological Chern number of $\nu = \pm 4$. Right: Dzyaloshinskii-Moriya, $D$ model. For any choice of the DM couplings, a spin disordered phase is stabilized and a continuous transition to the respective classical magnetic state is observed (arrow). 
    }
	\label{fig:pdmmft}
\end{figure}

\begin{figure}
	\centering
	\includegraphics[width=0.5\textwidth,clip]{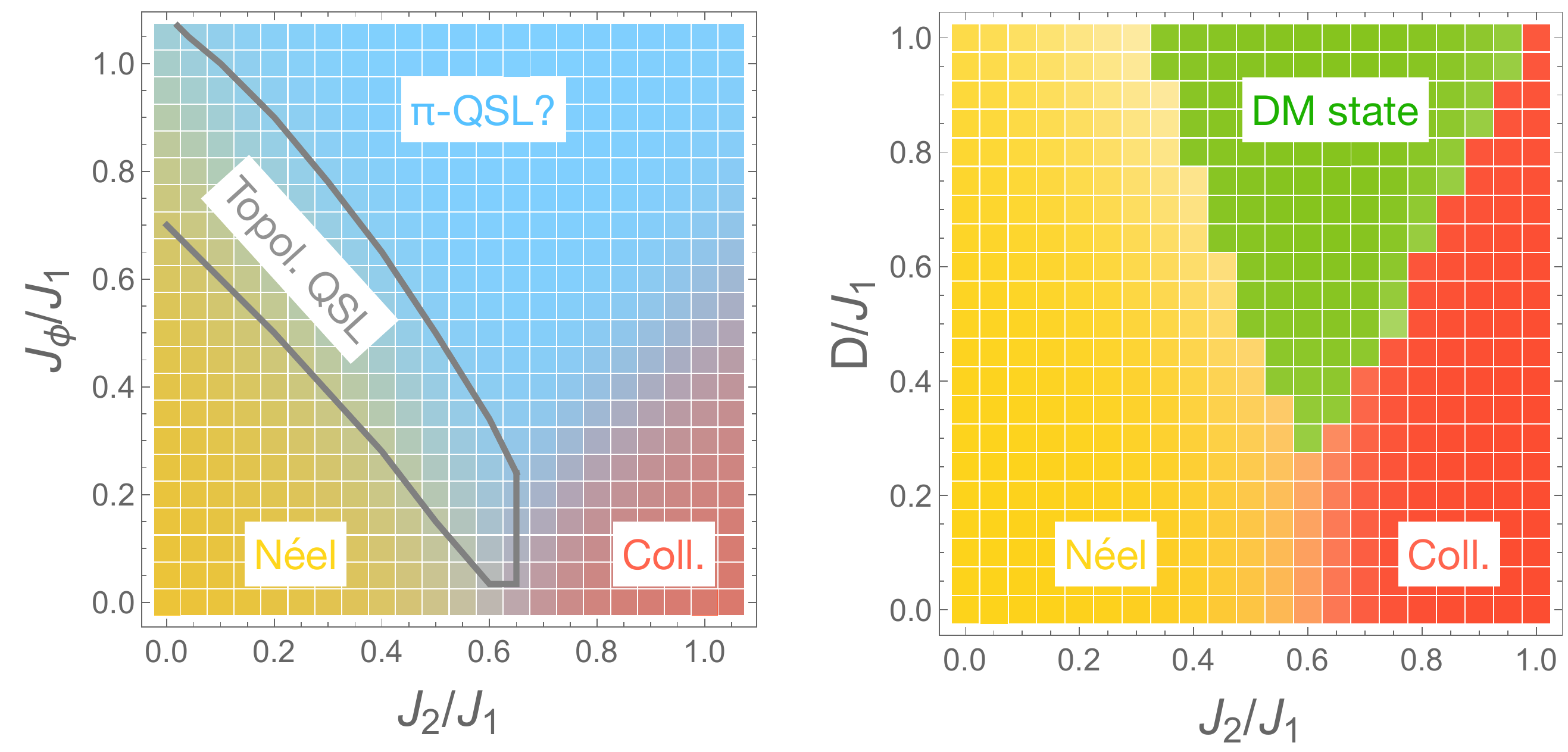}
\caption{Phase diagrams of the $J_\phi$ and $D$ models (illustrated for YBCO) obtained from ED calculations on a $4 \times 4$ cluster based on a quantum fidelity analysis. The more intense the color, the closest to 1 the fidelity. Three reference states are used, (i) the pure N\'eel at point $J_2=J_\phi=D=0$ (yellow), (ii) the collinear state $J_2=1,J_\phi=D=0$ (red) and (iii) the 
intermediate phase (blue) at $J_2=0.6, J_\phi=D=1.0$.  Left: orbital magnetic field, $J_\phi$ model presenting an intermediate topological QSL. Right: Dzyaloshinskii-Moriya, $D$ model and its first order transition to a DM state driven by $D$.
 }
	\label{fig:edpd}
\end{figure}  

{\em Orbital magnetic field and spin-orbit coupling effects.}
We have performed fully self-consistent MMFT calculations on the 
$J_1$-$J_2$-$J_\phi$ model(see Supplemental Material [\onlinecite{suppl}]). 
The resulting phase diagram shown in Fig.\ref{fig:pdmmft} (a) 
displays N\'eel, stripe and spin disordered 
phases. For $J_\phi=0$ we qualitatively recover the phase diagram 
of the $J_1$-$J_2$ Heisenberg model on the square lattice. 
Subject of an intense research activity since the discovery of 
cuprate superconductivity, recent state-of-the-art numerical works 
\cite{Sandvik2018,Becca2020,Poilblanc2020,Imada2021} find
a gapless quantum spin liquid and a valence bond solid 
around $J_2/J_1 \sim 0.5$. Our MMFT is partially consistent with 
this since it predicts a spin disordered valence bond crystal (VBC)
phase sandwiched between the magnetically ordered N\'eel and 
collinear phases. Note that, at the mean field level, all VBCs like the columnar or the staggered phases, as well as possible resonating placquette phases\cite{sizeofunitcell} are degenerate 
all consisting of disconnected singlets. On the other hand, under sufficient strong J$_\phi$, 
the MMFT finds that the $\pi$-QSL becomes the ground state  
in a broad region of the $J_2$-$J_\phi $ phase diagram.  
Although nonphysical large $J_\phi$ values are needed, 
it is well known that 
mean-field theory overestimates broken symmetry states. 
Indeed, we show below how quantum fluctuation effects
strongly suppress the critical $J_\phi$ at which the
$\pi$-QSL emerges in the phase diagram.

We now analyze the effect of the DM interaction parameterised by 
the ${\bf D}$ vectors(see Supplemental Material [\onlinecite{suppl}]). The DM is also
found to favor the gapped $\pi$-QSL (with Chern number $\nu = \pm 4$) but \textit{a contrario} to the orbital effect, it is never the absolute ground state of the $J_1$-$J_2$-$D$ model.
We have considered all possible MMFT self-consistent solutions  
analyzing in detail the competition between the $\pi$-QSL phase, the 
N\'eel and collinear magnetically ordered states and 
spin disordered states\cite{sizeofunitcell}
for five different DM vector choices [\onlinecite{dmchoices}]. We have found that the disordered states are always the ground state and that the effect of DM is to continuously deform both the GS and the QSL to eventually reach, when $D$ dominates, a magnetic order compatible with the considered $\textbf{D}_{ij}$. The typical resulting phase diagram is shown in Fig.\ref{fig:pdmmft}(b) for YBCO.
Interestingly, this phase diagram is qualitatively similar to the phase
diagram of the $J_1$-$J_2$-$J_\phi$ model discussed previously.
Our MMFT results suggest that even for very large DM, 
$D>0.5 J_1$, not present in the cuprates ($D<0.1 J_1$),  the $\pi$-QSL phase with an associated large thermal Hall effect is not the absolute ground state but still a robust and close excited state.
Nevertheless, it can still play an important role in effectively bringing the system closer to the $\pi$-QSL
phase. Thus, from our MMFT analysis we expect 
that, in the presence of the DM, the 
large thermal Hall effect could occur
at a smaller $J_\phi$ than for $D=0$. 

{\em Beyond Majorana mean-field theory.} 
An important question is whether the $\pi$-QSL can occur beyond the MMFT approach. 
In Fig.~\ref{fig:edpd} we show the quantum fidelity $|\langle \psi_\text{ref} | \psi_0 \rangle |$ where $| \psi_0 \rangle$ is the ground state for a given set of parameters 
and $ | \psi_\text{ref} \rangle $ is a reference state at known
limiting values of the parameters (see 
caption of Fig.~\ref{fig:edpd} for more details). 
For $J_\phi=D=0$, we identify the N\'eel and collinear phases a well as a possible disordered phase around $J_2 \sim 0.5-0.7$. We also find that the phases shown in Fig.~\ref{fig:edpd} arising between the Néel and collinear phase under $J_\phi$ or under $D$ are quite different. ED calculations of the spin structure factor(see Supplemental Material [\onlinecite{suppl}]) and spin chirality
$\langle T_{ijk} \rangle$ suggest that, while the intermediate phase is spin disordered (blue region) and chiral, $\langle T_{ijk} \rangle \neq 0 $, in the $J_\phi$ model, 
in the $D$-model it is non-chiral, $\langle T_{ijk} \rangle = 0$ and possibly also 
spin disordered as discussed below (green region). 
It however has to connect with the pure $D$ limit at which magnetic order 
induced by $D$ eventually sets in. 
Based on the consistent comparison of the ED and the MMFT phase diagrams of Fig.~\ref{fig:pdmmft}, we associate the (blue) spin disordered phase region with the MMFT $\pi$-QSL and the (green) magnetic DM with MMFT DM magnetically ordered phase.

 Within the intermediate spin disordered phase of the $J_\phi$-model, close to the Néel phase, we find a topological QSL. This QSL is chiral characterized by a non-zero Chern number, $\nu=1$, associated with a two-fold degenerate ground state well separated from 
 a continuum of excited states, as reported previously.\cite{Cirac2013} Interestingly this phase survives down to very low $J_\phi$ for $J_2 \sim 0.6$ as shown in Fig.~\ref{fig:edpd}. 
 In the $D$-model,
the evidence for a distinct and possibly spin disordered 
phase is corroborated by calculations combining the 
quantum fidelity, the gap and the spin structure factor. Indeed, in contrast to the orbital magnetic field case, the spin phase induced
by $D$ is clearly delimited by a gap closing associated with level crossings (see Supplemental Material [\onlinecite{suppl}]) suggesting, to the best of our knowledge, a novel intermediate phase induced by spin-orbit coupling. We let the analysis of this phase as a future work.

The MMFT could be used to unveil the 
nature of the chiral QSLs found in other numerical studies of the Heisenberg model on the triangular lattice with four-spin terms\cite{Moore2021} and with three-spin orbital magnetic field  terms\cite{Lauchli2017} and on spin models of Kagomé Mott insulators.\cite{Vidal2014,Messio2012,Sheng2014a,Sheng2014b}

{\it Connection with cuprate materials.}
In cuprates\cite{Pavarini2001} $t_1 \sim 0.45$ eV, $U \sim 8$ eV, 
$J_1 \sim 0.1$ eV, and $t_2 \sim 0.35 t_1$ (YBCO),  $t_2 \sim 0.15 t_1$ (LSCO) so that $J_2/J_1 \sim 0.12$ (YBCO), $J_2/J_1 \sim 0.023$ (LSCO). 
Since DM is only of a few meV, $D/J \lesssim 0.1$, we would predict 
that the system under no magnetic field 
is in the N\'eel phase as indeed is observed in undoped cuprates. 
If a magnetic field of $B \sim 10$ T is applied 
to the system, the flux term, $J_\phi/J_1 \sim 10^{-4}-10^{-3} $ would be tiny. Based on the phase diagrams obtained here,
in such parameter range appropriate for cuprate materials, $ J_2/J_1 \sim 0.1$ and $J_\phi \rightarrow 0$, 
the system would be immersed in the
N\'eel phase and so no thermal Hall effect  
would be expected based on our analysis. 
However, since the MMFT finds (see Supplementary Material [\onlinecite{suppl}]) that the $\pi$-QSL can coexist with the N\'eel AF, the $\pi$-QSL amplitude  
in the GS can still lead to a large thermal Hall effect. Such hybrid $\pi$-QSL + N\'eel AF state 
may also account for the anomaly observed around $(\pi,0)$ 
in the magnon dispersion of square lattice antiferromagnets\cite{Ivanov2015}
which could result from teh decay of 
$S=1$ magnons into a continuum of Majorana excitations 
associated with the $\pi$-QSL.

{\it Conclusions.}
We report a novel chiral QSL state with Majorana excitations and Chern number $\nu= \pm 4$ 
which occurs as an excited state of the Heisenberg model on the square lattice. The thermal Hall conductivity of such Majorana $\pi$-QSL state leads to large absolute values $|\kappa_{xy}/T| \sim  (k_B^2/\hbar)$ consistent with experiments in Mott insulating cuprates.  
MMFT predicts that this Majorana 
$\pi$-QSL becomes the ground state under 
sufficiently large 
$J_\phi$ consistent with a topological 
chiral QSL found in ED. The DM present 
in cuprates plays a secondary role 
eventually suppressing critical $J_\phi$'s towards more physically realistic values. 
The large thermal Hall effect observed in 
Mott insulating cuprates can be interpreted 
from their proximity to a Majorana $\pi$-QSL transition which can be triggered by an external magnetic field via its orbital effect.

\acknowledgments
J. M. acknowledges financial support from (RTI2018-098452-B-I00) MINECO/FEDER, Uni\'on Europea and the Mar\'ia de Maeztu Program for Units of Excellence in R\&D (Grant No. CEX2018- 000805-M).

\appendix


\begin{thebibliography}{30}
\bibitem{dmchoices}{The corresponding DM vectors for the 5 cases considered in this work are given by: YBCO with $\vec{d}_1=\vec{d}_3=(d,0,0)$ and $\vec{d}_2=\vec{d}_4=(0,-d,0)$, LSCO-LTO with $\vec{d}_1=-\vec{d}_3=(d,0,0)$ and $\vec{d}_2=-\vec{d}_4=(0,-d,0)$, LSCO-LTT with $\vec{d}_1=-\vec{d}_3=(0,d,0)$ and $\vec{d}_2=-\vec{d}_4=(0,d,0)$, \textit{Rashba}  with $\vec{d}_1= \vec{d}_3=(0,d,0)$ and $\vec{d}_2= \vec{d}_4=(-d,0,0)$ and \textit{Dresselhaus}  with $\vec{d}_1= \vec{d}_3=(0,-d,0)$ and $\vec{d}_2= \vec{d}_4=(-d,0,0)$ }
\end{thebibliography}
\end{document}


\title{Supplementary material for Majorana chiral spin liquid in Mott insulating cuprates}
  \author{Jaime Merino}
\affiliation{Departamento de F\'isica Te\'orica de la Materia Condensada, Condensed Matter Physics Center (IFIMAC) and
Instituto Nicol\'as Cabrera, Universidad Aut\'onoma de Madrid, Madrid 28049, Spain}
\author{Arnaud Ralko}
\affiliation{Institut N\'eel, UPR2940, Universit\'e Grenoble Alpes et CNRS, Grenoble 38042, France}
 \date{\today}
\maketitle

\section{Majorana mean-field theory of the Heisenberg model on square lattice}
We first provide the details of the Majorana mean-field theory (MMFT) performed on the $J_1-J_2$ Heisenberg model on a square lattice. We introduce a Majorana representation of the spin operators \cite{kitaev2006}: 
\begin{equation}
S^\alpha_i= {i \over 2} b^\alpha_i c_i,
\end{equation}
where $\alpha=x,y,z$. Although the Hilbert space is 
enlarged by the Majorana representation, the actual 
physical space can be recovered by imposing 
the set of constraints over the Majorana fermions: 
\begin{eqnarray}
b^z_i c_i= b^y_i b^x_i,
\nonumber \\
b^y_i c_i = b^x_i b^z_i,
\nonumber \\
b^x_i c_i = b^z_i b^y_i,
\label{eq:constraints}
\end{eqnarray}
corresponding to a single occupancy constraint on the original fermions.

Any bilinear of spins $S^\alpha_i S^\beta_j$ can be mean-field Hartree-Fock decoupled as:
\begin{eqnarray}
\label{decoupling}
&+&\frac{1}{4} \left[ \langle b_i^\alpha  i c_i \rangle  b_j^\beta i c_j + \langle b_j^\beta i c_j \rangle b^\alpha_i i c_i -\langle b_i^\alpha i c_i \rangle \langle  b_j^\beta i c_j\rangle \right] \nonumber \\
&-&\frac{1}{4} \left[ \langle b_i^\alpha i b_j^\beta \rangle   c_i i c_ j + \langle c_i  i c_j \rangle b^\alpha_i i b^\beta_j - \langle b^\alpha_i i b^\beta_j \rangle \langle c_i i c_j \rangle \right]
\nonumber \\
&+&\frac{1}{4} \left[ \langle b_i^\alpha  i c_j \rangle  b_j^\beta i c_i + \langle b_j^\beta i c_i \rangle b^\alpha_i i c_j -\langle b_i^\alpha i c_j \rangle \langle  b_j^\beta i c_i\rangle \right] 
\end{eqnarray}
where the first three terms are associated with magnetic ordering while the rest with spin liquid formation.  
This decoupling is performed on the Heisenberg terms of the Hamiltonian ($\alpha = \beta$).The constraints in Eq. (\ref{eq:constraints})
are enforced at the mean-field level through 
the set of Lagrange multipliers, $\{ \lambda_i \}$. Since there 
is no contribution from the last three terms we neglect them in 
our present analysis.

In order to be able to describe collinear phases four-site unit cells are used in actual calculations. Here, we describe the implementation of the MMFT equations assuming a two-site unit cell allowing for
N\'eel and/or $\pi$-flux QSL states. The diagonalization
of the hamiltonian is performed in momentum space 
by first performing a Fourier transform of the Majoranas:
\begin{equation}
c_j^a=\sqrt{2 \over N_c} \sum_{{\bf k}} e^{i {\bf k} {\bf R}_j} c^a({\bf k}),  
\end{equation}
where we have used here, for convenience, the notation: 
$c_j^0=c_j, c_j^1=b_j^x, c_j^2=b_j^y, c_j^3=b_j^z$.
On the other hand, the operators in momentum space, 
$c^a({\bf k})$, satisfy 
$c^a({\bf k})=c^{a\dagger}(-{\bf k})$ due to the
property: $c^a_i=c^{a\dagger}_i$  
which leads to standard fermionic anticommutation relations: 
$\{ c^{a\dagger}({\bf k}), c^{b\dagger}({\bf k'}) \} =\{ c^a({\bf k}), c^b({\bf k'}) \} =  0,\{ c^a({\bf k}), c^{b\dagger}({\bf k'}) \} = 
\delta_{ab} \delta({\bf k}-{\bf k'}) $.
We have:
\begin{equation*}
H^{MMF}({\bf k})=
\begin{pmatrix}
H_{AA} ({\bf k }) & H_{AB} ({\bf k} )   \\
H_{BA}({\bf k})  & H_{BB}({\bf k} ) \\
\end{pmatrix}
\end{equation*}
where $A,B$ are the two sublattices in which the original lattice is divided. Since the spin operators are expressed through
four Majorana fermions $\{c, b^x, b^y, b^z \}$, each block, $H_{ab}$ with 
$a, b= A, B$ is a $4 \times 4$ matrix.

The hamiltonian block associated with sites in the same sublattice, $a=A,B$, reads:
\begin{equation*}
H_{aa}({\bf k})=
\begin{pmatrix}
H^{00}_{aa}({\bf k})   &   -{i \over 4} \lambda_{ax} & -{i \over 4} \lambda_{ay} &  -{i \over 4} \lambda_{az} \\
{i \over 4} \lambda_{ax} &  H^{xx}_{aa}({\bf k}) &  {i \over 4} \lambda_{az}& -{i \over 4} \lambda_{ay} \\
{i \over 4} \lambda_{ay} &  -{i \over 4} \lambda_{az} &  H^{yy}_{aa}({\bf k})  & {i \over 4} \lambda_{ax} \\
{i \over 4} \lambda_{az}  &  {i \over 4} \lambda_{ay}  & -{i \over 4}\lambda_{ax} & H^{zz}_{aa}({\bf k})  \\
\end{pmatrix}
\end{equation*}
where:
\begin{eqnarray}
H^{00}_{aa}({\bf k}) &=& -{iJ_2 \over 4}\sum_{\langle \langle i_a j_a \rangle \rangle } \sum_\alpha \langle i b_{i_a}^\alpha b_{j_a}^\alpha \rangle e^{i {\bf k} \cdot {\bf \delta}_{i_aj_a}} 
\nonumber \\
H^{\alpha\alpha}_{aa}({\bf k}) &=& -{iJ_2 \over 4}\sum_{\langle \langle i j \rangle \rangle }   \langle i c_{i_a}  c_{j_a} \rangle e^{i {\bf k} \cdot {\bf \delta}_{i_aj_a} }.
\end{eqnarray}
with: $\alpha=x, y, z$.

The hamiltonian block associated with the $A - B$ interaction reads:
\begin{equation*}
H_{AB}({\bf k})=
\begin{pmatrix}
H^{00}_{AB}({\bf k})   &   0 & 0 &  0 \\
0 &  H^{xx}_{AB}({\bf k}) &  0 & 0 \\
0 &  0 & H^{yy}_{AB}({\bf k})  & 0 \\
0 &  0 & 0 & H^{zz}_{AB}({\bf k})  \\
\end{pmatrix}
\end{equation*}
where:
\begin{eqnarray}
H^{00}_{AB}({\bf k}) &=& -{iJ_1 \over 4} \sum_{\langle i_Aj_B \rangle }\sum_\alpha \langle i b^\alpha_{i_A} b^\alpha_{j_B} \rangle e^{i {\bf k} \cdot {\bf \delta}_{i_Aj_B}} 
\nonumber \\
H^{\alpha\alpha}_{AB}({\bf k}) &=&-{iJ_1 \over 4} \sum_{\langle i_Aj_B \rangle } \langle i c^\alpha_{i_A} c^{\alpha}_{j_B} \rangle e^{i {\bf k}\cdot {\bf \delta}_{i_Aj_B}}.
\end{eqnarray}

The self-consistent loop proceeds as follows. 
A random guess of variational parameters, $\langle i c^\alpha_{i_a} c^{\alpha}_{j_b} \rangle, \langle i b^\alpha_{i_a} b^\alpha_{j_b} \rangle$ 
is injected in $H^{MMF}({\bf k})$ and the hamiltonian diagonalized.  
The set of Lagrange multipliers $\{ \lambda_i \}$ is fixed by the single occupancy constraint in Eq. (\ref{eq:constraints}) using a least square minimization. 
Finally, we find a new set of variational parameters obtained from the 
mean-field ground state which is a Slater determinant constructed
out from the filled orbitals of the mean-field 
hamiltonian. These three steps are repeated until 
convergence with the desired tolerance is reached.

We now analyze the $\pi$-flux solution 
to the $J_1$-$J_2$ Heisenberg model discussing the
role played by N\'eel antiferromagnetism. We also 
explore other possible quantum spin disordered solutions 
such as the valence bond crystal (VBC). We focus on 
the $0< J_2 0.5$ regime which is sufficient for 
our purposes.




\subsection{$\pi$-QSL solution}

For the $\pi$-flux QSL bond pattern shown in Fig. 1 of the main text, the Majorana mean-field 
hamiltonian is:
\begin{eqnarray}
H^{00}_{AA}({\bf k}) &=& J_2\sum_\alpha \langle b_{i_A}^\alpha i b_{j_A}^\alpha \rangle \cos(k_x) \sin(k_y)
\nonumber \\
H^{\alpha\alpha}_{AA}({\bf k}) &=& J_2 \langle c_{i_A } i c_{j_A} \rangle \cos(k_x) \sin(k_y),
\nonumber \\
H_{AB}^{00} ({\bf k} ) &=& i {J_1 \over 4}\sum_{\alpha}  \langle b^\alpha_{iA} i b^\alpha_{jB} \rangle  (-2 i \sin(k_x) + 2 \cos(k_y) )
\nonumber \\
H_{AB}^{\alpha\alpha} ({\bf k}) &=& i {J_1 \over 4}  \langle  c_{iA} i c_{jA} \rangle  
(-2 i \sin(k_x)+ 2\cos(k_y) ).
\end{eqnarray}
where: $\langle c_{i_A} i c_{j_B} \rangle = \langle b^\alpha_{i_A} i b^\alpha_{j_B }\rangle > 0 $,
and $\langle b_{i_A}^\alpha i b_{j_A}^\alpha \rangle=\langle c_{i_A } i c_{j_A} \rangle<0$ with  $\alpha=x,y, z$
for the flux pattern in Fig. 1 of the main text
which has a Chern number of $\nu=-4$.
Also we have: $H^{00}_{BB}({\bf k})=-H^{00}_{AA}({\bf k}) $ and $H^{\alpha\alpha}_{BB}({\bf k})=-H^{\alpha\alpha}_{AA}({\bf k}) $. 

The resulting Majorana dispersions can be obtained analytically: $\pm  3 \gamma({\bf k}), \pm \gamma({\bf k})\text{ (triply degenerate)}$, with:
\begin{widetext}
\begin{equation}
\gamma({\bf k})= {J_1 \over 2} \sqrt{  c_1^2 ( \sin^2(k_x) +\cos^2(k_y))  
+  ({J_2 \over J_1})^2 ( 2 c_2 \cos(k_x) \sin(k_y))^2 },
\label{eq:gamma}
\end{equation}
\end{widetext}
where: $ c_1=\langle c_{iA} i c_{j B} \rangle= \langle b_{i_A}^\alpha i b_{j_B}^\alpha \rangle, 
 c_2=\langle c_{i_A} i c_{j_A} \rangle=\langle b_{i_A}^\alpha i b_{j_A}^\alpha \rangle $. These are the expressions quoted in
 the main text.

The dependence of the $\pi$-QSL energy with $J_2/J_1$ is shown in Fig. \ref{fig:hybrid}.The energy 
is almost constant in the range shown: $E(J_2=0)=-0.3442, E(J_2=0.5)=-0.3443$. The lowering
of the energy with $J_2$ due to the opening of the topological gap is evident, for instance, 
$E(J_2=1)=-0.372675$.

\subsection{$\pi$-QSL + N\'eel antiferromagnetism}

\begin{figure}
	\centering
	\includegraphics[width=4cm,clip]{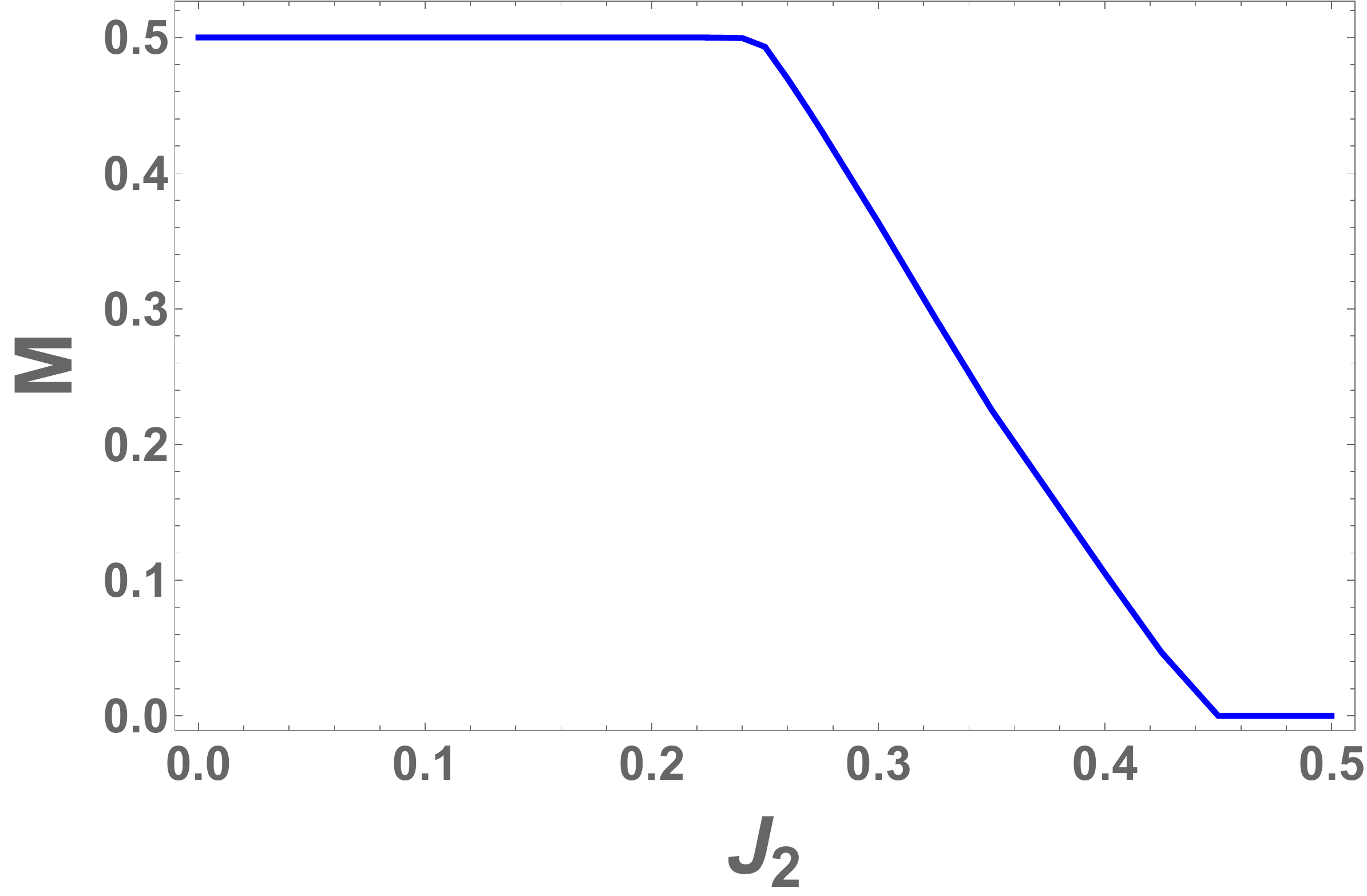}
	\includegraphics[width=4cm,clip]{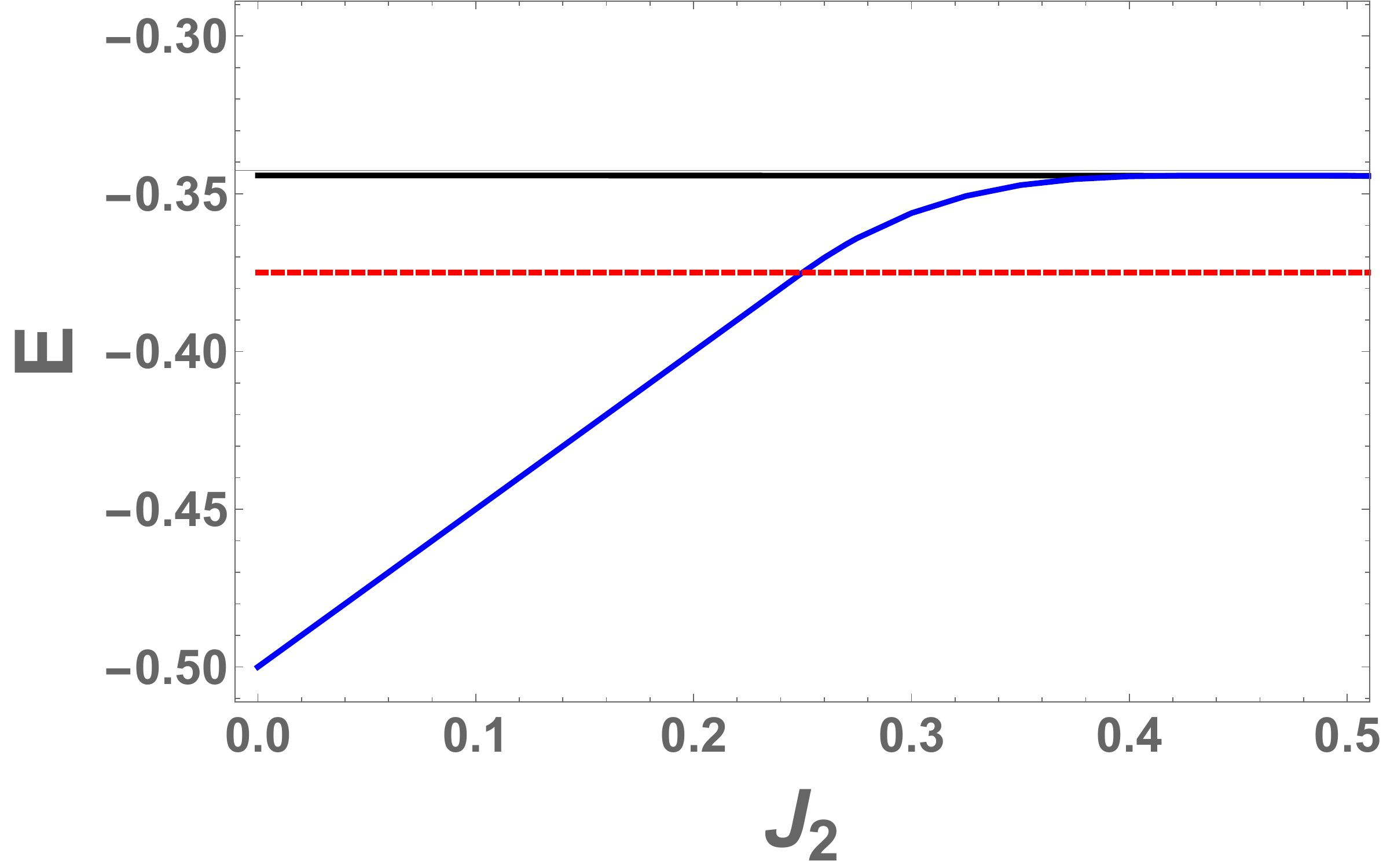}
	\includegraphics[width=4cm,clip]{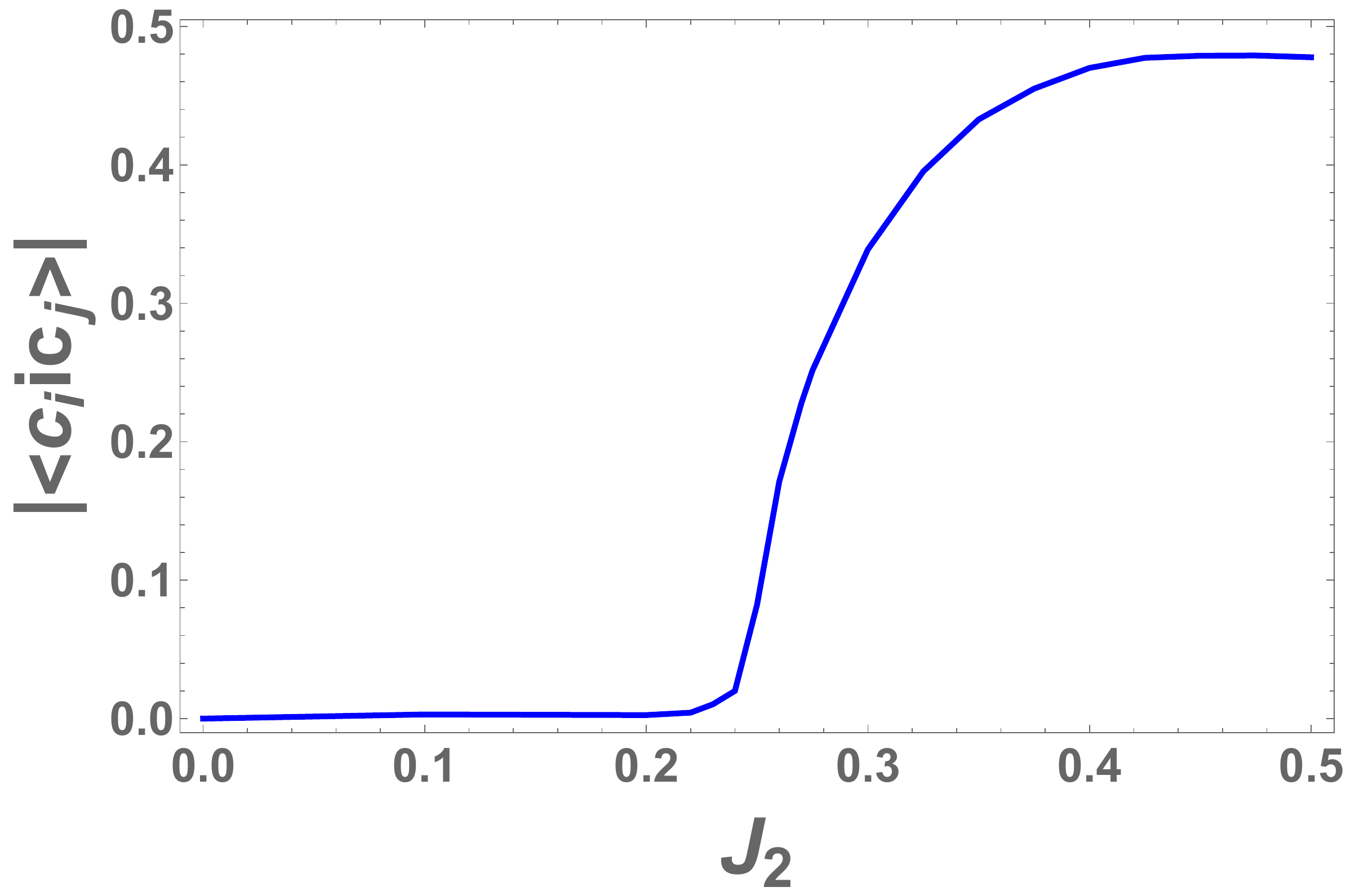}
	\includegraphics[width=4cm,clip]{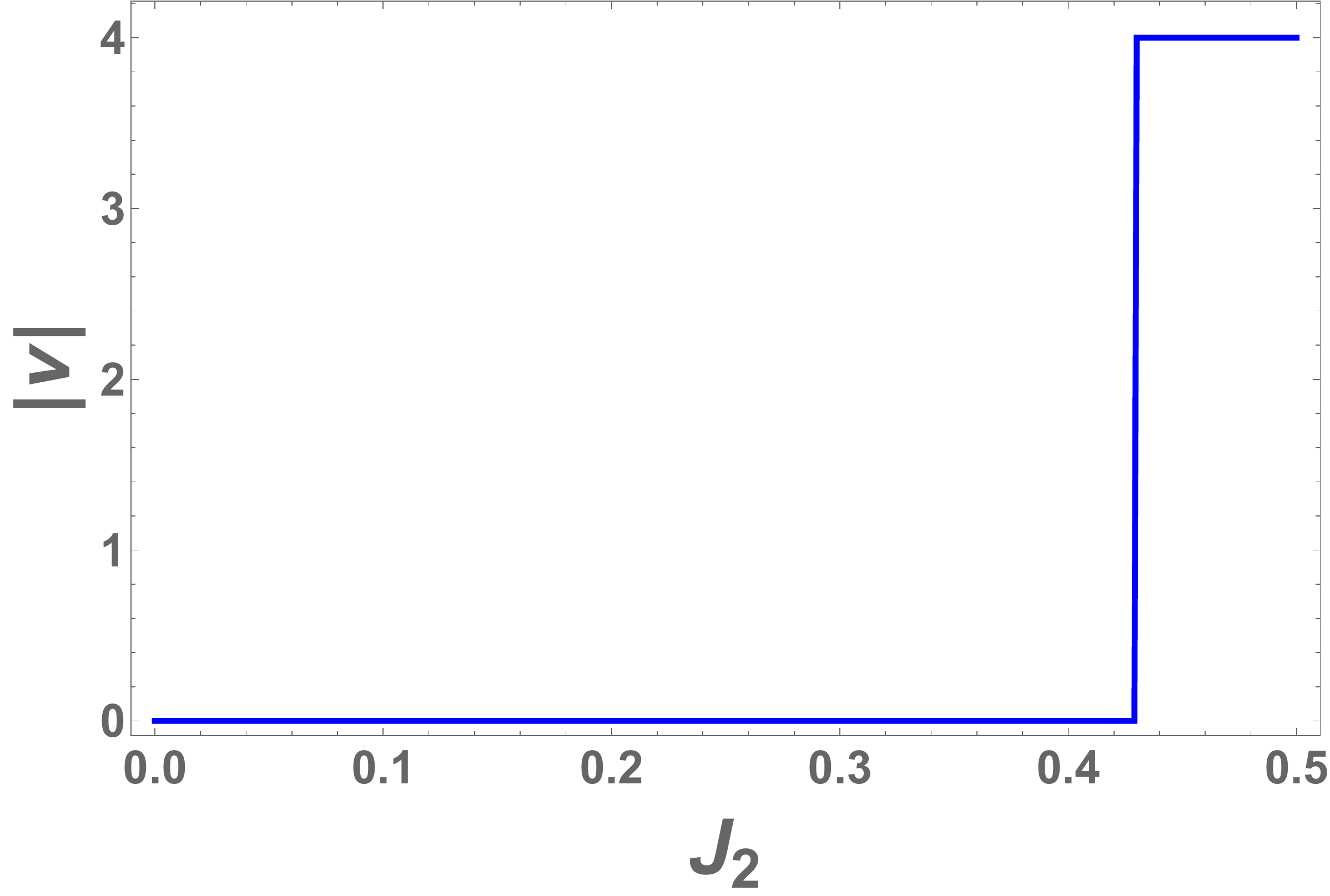}
	\caption{Dependence of properties of various MMFT solutions with $J_2$.  From top to bottom: 
	antiferromagnetic moment (M), energy per site (E)
	absolute value of $\pi$-QSL amplitude ($|\langle c_i i c_j \rangle |)$) and Chern number ($|\nu|$) are shown fthe hybrid $\pi$-QSL (blue solid lines). 
	The energies of the spin disordered VBC (red solid line) and the pure ($M=0$) $\pi$-QSL (black solid line) are also shown for comparison. 
	$J_1=1$ in this plot.} 
	\label{fig:hybrid}
\end{figure}

An important aspect of the physics of undoped cuprates though,
 which we have ignored up to now, is N\'eel antiferromagnetism (AF). 
 For $J_2=0$ we expect a N\'eel state in the square lattice. We have searched for 
 hybrid $\pi$-flux QSL +  N\'eel AF solutions. Indeed a non-zero  AF moment is present for $0 < J_2 < 0.43$
 as shown in Fig. \ref{fig:hybrid}.  For $J_2<0.25$ the N\'eel-AF is fully saturated, $M=0,5$, 
 and so the energy follows the classical linear dependence with $J_2$: 
 $E_{Neel}=-0.5+ 0.5J_2$  (taking $J_1=1$) shown in Fig. \ref{fig:hybrid}. 
 In contrast, for $0.24< J_2 < 0.43$, a $\pi$-QSL+N\'eel AF {\it i. e.} a state having  $M <0.5$ and 
 $|\langle c_ i c_j \rangle| \neq 0$ arises. A topological transition occurs at $J_2 \sim 0.43$ 
where N\'eel antiferromagnetism fades away and the 
pure $\pi$-QSL with $|\nu|=4$ wins. Fig. \ref{fig:hybrid}
indeed shows how the energy of the $\pi$-QSL+N\'eel AF converges 
to the energy of the pure $\pi$-QSL a $J_2/J_1$ at sufficiently large 
$J_2/J_1$.

Although our MMFT $\pi$-QSL + N\'eel AF is always gapped,
the gap for $J_2 < 0.43$ associated with the N\'eel order is trivial while the gap for $J_2 >0.43$ in the $\pi$-QSL phase is topological 
with Chern number, $|\nu|=4$. Hence, even though a hybrid $\pi$-flux QSL + N\'eel AF actually occurs in the MMFT equations, such state is non-topological, in fact only the pure $\pi$-QSL is topological. However, this could be an artifact of the MMFT in which broken symmetry states are treated classically.
 
 The hybrid  $\pi$-QSL + N\'eel AF state is not the lowest energy 
 state in the whole $0 < J_2/J_1< 0.5$ range. The spin disordered valence bond crystal (VBC) 
 consisting of singlets between n.n. spins, has an energy per site of 
 $E_{VBC}=-0.375$ for any $J_2/J_1$ since
 the singlets are effectively decoupled.
 As shown in Fig. \ref{fig:hybrid} the VBC state wins for $J_2/J_1 >0.25$, actually the onset of the hybrid $\pi$-QSL + N\'eel AF state occurs.  Although our total energy analysis suggests that a 
 direct transition from a N\'eel AF to a VBC occurs around 
 $J_2 \sim 0.25 J_1$, it also indicates that a $\pi$-QSL +
 N\'eel AF is a possible self-consistent solution of the MMFT. Theories beyond the Majorana mean-field treatment including quantum fluctuations may stabilize a coexistent $\pi$-QSL + N\'eel AF state 
 with non-trivial topological properties.

\section{Three-spin magnetic orbital term}

We now apply the MMFT to the magnetic orbital contribution considered
in the main text:
\begin{equation}
H_\phi=J_\phi \sum_{\triangle} {\bf S}_i \cdot ({\bf S}_j  \times {\bf S}_k),
\end{equation}
where the sum is taken over all triangular placquettes of the square lattice.

This kind of three-spin terms has been discussed in the context of bosonic and fermionic spinon theories  
of the Hubbard model in the large-$U$ limit. \cite{Motrunich2006,Lee2019,Sachdev2019a} Actually fermonic mean-field theories 
provide an exact solution of the large-$N$ limit of the model as has been shown in 
the SU(N) Heisenberg model.\cite{Auerbach}
In order to make direct contact with these approaches we apply the MMFT decoupling directly 
onto the large-$N$ expression for the chiral three-spin terms. Our starting point is the full expression of the 
three-spin contribution expressed in terms of Abrikosov fermions\cite{Motrunich2006}:
\begin{equation}
{\bf S}_i \cdot ({\bf S}_j \times {\bf S}_k)=-{i \over 4} (P_{ijk} -P^\dagger_{ijk} )
\end{equation}
where $P_{ijk}=(f^\dagger_{i\alpha} f_{i\beta})  (f^\dagger_{j \beta} f_{j \gamma}) (f^\dagger_{k\gamma} f_{k\alpha}) $ exchanges three fermions
around the triangular placquette. Note that Einstein notation for the sums over spin indices is assumed here. 
The mean-field treatment of this contribution consists on contracting only the fermionic 
operators with the same spin which leads to the large-$N$ expression:
\begin{eqnarray}
{\bf S}_i \cdot ({\bf S}_j \times {\bf S}_k)  &=& -{i \over 4} [ (f^\dagger_{j \beta} f_{i \beta})(f^\dagger_{k \gamma} f_{j \gamma}) (f^\dagger_{i\alpha} f_{k\alpha})
\nonumber \\
&-& (f^\dagger_{k\alpha} f_{i\alpha})  (f^\dagger_{j \gamma} f_{k \gamma}) (f^\dagger_{i\beta} f_{j\beta}) ].
\end{eqnarray}
The fermions satisfy: $\sum_\alpha f^\dagger_{i\alpha} f_{i\alpha}=1$. It can be shown that even for the not so large value $N=2$ the 
contractions neglected do not change qualitatively the physics. Defining $\chi_{ij}= f^\dagger_{j\alpha} f_{i\alpha}  $  the mean-field
hamiltonian, $H_\phi$ reads:
\begin{eqnarray}
H^{MF}_{\phi}&=& i {J_\phi \over 4} \sum_{\triangle} [ \chi_{ij} \langle \chi_{jk}  \rangle \langle \chi_{ki} \rangle  -\langle \chi_{ik} \rangle \langle \chi_{kj} \rangle \chi_{ji} \nonumber \\
&+& \langle \chi_{ij} \rangle  \chi_{jk}   \langle \chi_{ki} \rangle
- \langle \chi_{ik} \rangle  \chi_{kj} \langle \chi_{ji} \rangle
\nonumber \\
&+&\langle \chi_{ij}  \rangle \langle \chi_{jk}  \rangle  \chi_{ki}  -  \chi_{ik}  \langle \chi_{kj} \rangle \langle  \chi_{ji}  \rangle \nonumber \\
 &-& 2 ( \langle \chi_{ij}  \rangle \langle \chi_{jk}  \rangle \langle \chi_{ki} \rangle  -\langle \chi_{ik} \rangle \langle \chi_{kj} \rangle \langle \chi_{ji} \rangle ) ],
\nonumber \\
\end{eqnarray}
where the sum is over all triangular placquettes with vertices $ijk$. 

Performing a Majorana transformation: 
\begin{eqnarray}
f_{i\uparrow} &=& {1 \over 2} (c_i-i b_i^z), f^\dagger_{i \uparrow} = {1 \over 2} (c_ i + i b_i^z)  
\nonumber \\
f_{i\downarrow} &=& {1 \over 2} (b_i^y -i b_i ^x) , f^\dagger_{i \downarrow} = {1 \over 2} (b_i^y + i b_ i^x),
\end{eqnarray}
leads to the decomposition in terms of the Majoranas:
\begin{widetext}
\begin{equation}
f^\dagger_{i\alpha} f_{j\alpha}= {1 \over 4} ( c_i c_j -i c_i b_j^z + i b_i^z c_j +b_i^z b_j^z +b_i^y b_j ^y-i b_i^y b_j^x + i b_i^x b_j ^y +b_i^x b_j^x ).
\end{equation}
\end{widetext}

We consider mean-field solutions with non-zero bond fluxes ${\it i. e.}$ $\langle \chi_{ij}  \rangle = \pm i R_{ij} $ with $R_{ij} \in  \mathbb{R} $ leading to the terms 
\begin{widetext}
\begin{eqnarray}
\langle \chi_{jk} \rangle \langle \chi _{ki} \rangle f^\dagger_{i\alpha} f_{j\alpha} - \langle \chi _{ik} \rangle \langle \chi_{kj} \rangle  f^\dagger_{j\alpha}f_{i\alpha} &=& 
\pm {1 \over 4}  R_{jk} R_{ki}  \nonumber \\
&\times& [ ( c_i c_j -c_j c_i) + 
( b^x_i b^x_j -b^x_j b^x_i) + ( b^y_i b^y_j -b^y_j b^y_i) +( b^z_i b^z_j -b^z_j b^z_i)  ].
\end{eqnarray}
\end{widetext}
where the final overall sign in front depends on the final bond flux orientation. Hence, the final hamiltonian, $H^{MF}_\phi$,
will only contain diagonal contributions. Since $H^{MF}_\phi$ as well
as the Heisenberg hamiltonian is diagonal in the Majoranas, we have
that:
\begin{equation}
R_{ij}={1 \over 4} ( \langle c_i c_j \rangle + \sum_\alpha 
\langle b_i^\alpha b_j^\alpha \rangle ).
\end{equation}

\begin{figure}
	\centering
	\includegraphics[width=4cm,clip]{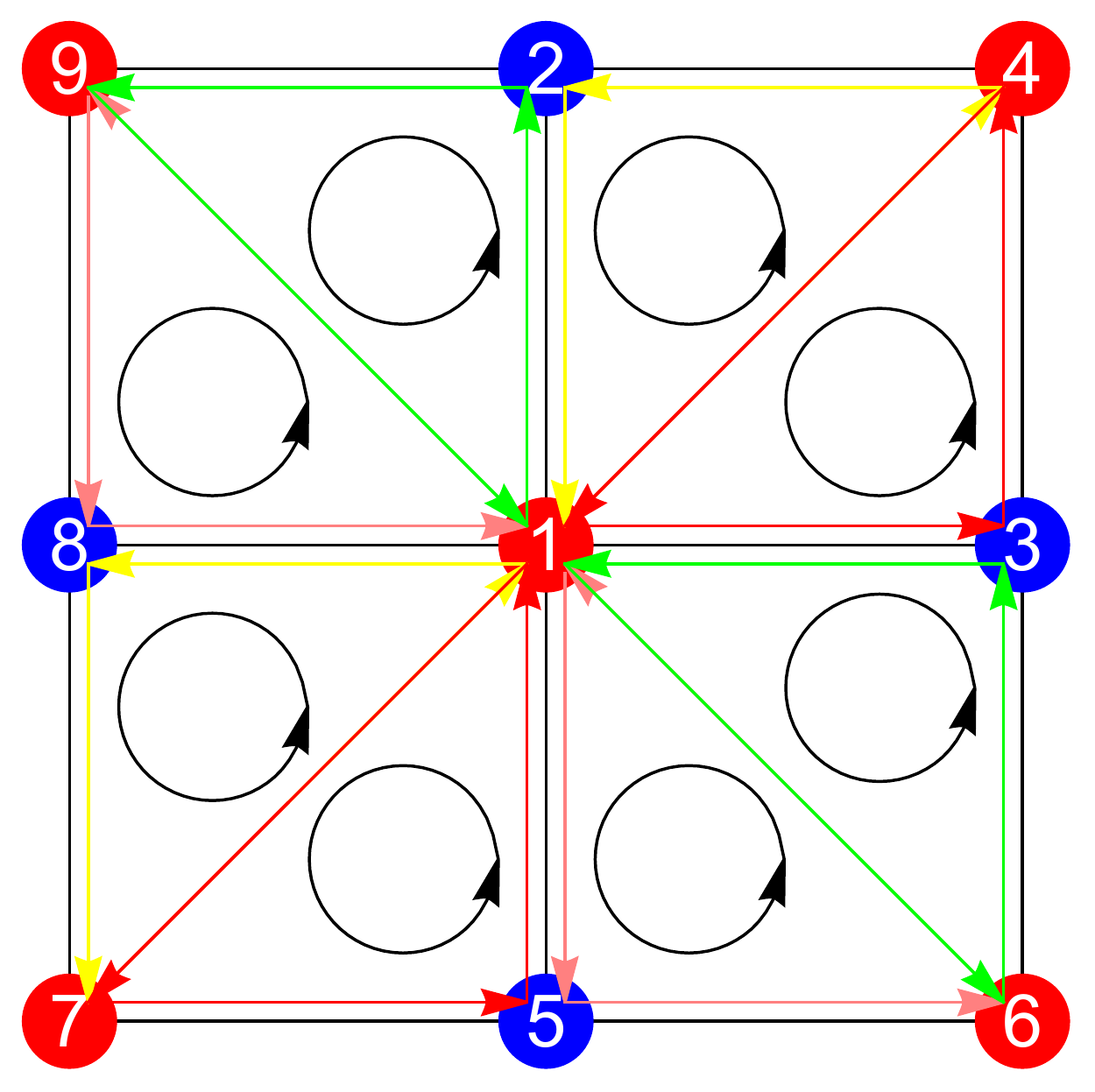}
	\caption{Three-spin magnetic orbital interactions. } 
	\label{fig:fluxterms}
\end{figure}  

The eight three-spin placquettes renormalising the 
Heisenberg exchange couplings between an $A$ spin   
and its nearest-neighbor $B$ and also its next-nearest-neighbor 
$B$ spin are shown in Fig. \ref{fig:fluxterms}.
While the three-spin terms:
\begin{equation}
   {\bf S}_1 \cdot ({\bf S}_4 \times {\bf S}_2 + {\bf S}_3 \times {\bf S}_4 + 
   {\bf S}_8 \times {\bf S}_7 + {\bf S}_7 \times {\bf S}_5  ) 
\end{equation}
are involved in the $1-4$ and $1-7$ n.n.n. couplings, the terms: 
\begin{equation}
   {\bf S}_1 \cdot ({\bf S}_6 \times {\bf S}_3 + {\bf S}_5 \times {\bf S}_6 + 
   {\bf S}_2 \times {\bf S}_9 + {\bf S}_9 \times {\bf S}_8  ) 
\end{equation}
renormalize the n.n.n. $1-6$ and $1-9$ exchange 
couplings. 
On the other hand the terms:
\begin{widetext}
\begin{equation}
   {\bf S}_1 \cdot ({\bf S}_3 \times {\bf S}_2 + {\bf S}_3 \times {\bf S}_4 + {\bf S}_6 \times {\bf S}_3 + {\bf S}_5 \times {\bf S}_3  +
  {\bf S}_2 \times {\bf S}_8 + {\bf S}_9 \times {\bf S}_8 + 
  {\bf S}_8 \times {\bf S}_7 
  \nonumber  \\
  + {\bf S}_8 \times {\bf S}_5 ) 
\end{equation}
\end{widetext}
renormalize the n.n. $1-3$ and $1-8$ couplings while:
\begin{widetext}
\begin{equation}
   {\bf S}_1 \cdot ({\bf S}_2 \times {\bf S}_9 + {\bf S}_2 \times {\bf S}_8 + {\bf S}_4 \times {\bf S}_2 + {\bf S}_3 \times {\bf S}_2
 + {\bf S}_5 \times {\bf S}_6 + {\bf S}_5 \times {\bf S}_3 + {\bf S}_7 \times {\bf S}_5 + {\bf S}_8 \times {\bf S}_5) ,  
\end{equation}
\end{widetext}
the n.n. $1-2$ and $1-5$ couplings.

The three-spin terms evaluated on zero-flux mean-field solutions 
such as a zero-flux QSL or a VBC solution
leads to cancellations so that $ \langle H^{MF}_\chi=0 \rangle$. Hence, to illustrate
the effect of the three-spin terms 
we evaluate $H^{MF}_\chi$ on a $\pi$-QSL bond pattern. Introducing the 
corresponding MMFT bond amplitudes $\langle c_i i c_j \rangle$ 
and $\langle b^\alpha_i i b^\alpha_j \rangle$  
in $R_{ij}$ we find the complete mean-field 
hamiltonian (including the Heisenberg contribution):
\begin{widetext}
\begin{eqnarray}
H^{00}_{AA}({\bf k}) &=& (J_2 \sum_\alpha \langle b_{i_A}^\alpha i b_{j_A}^\alpha \rangle + J_\phi R_1^2) \cos(k_x) \sin(k_y)
\nonumber \\
H^{\alpha\alpha}_{AA}({\bf k}) &=& (J_2 \langle c_{i_A } i c_{j_A} \rangle + J_\phi R_1^2)  \cos(k_x) \sin(k_y),
\nonumber \\
H_{AB}^{00} ({\bf k} ) &=& {i \over 4} ( J_1 \sum_{\alpha} \langle b^\alpha_{iA} i b^\alpha_{jB} \rangle + J_\phi R_1 R_2 )  (-2 i \sin(k_x) + 2 \cos(k_y) )
\nonumber \\
H_{AB}^{\alpha\alpha} ({\bf k}) &=& {i \over 4} ( J_1 \langle  c_{iA} i c_{jA} \rangle + J_\phi R_1 R_2 ) (-2 i \sin(k_x)+ 2\cos(k_y) ),
\label{eq:hamflux}
\end{eqnarray}
\end{widetext}
where $R_{ij}=R_1$ for nearest-neighbor and $R_{ij}=R_2$ for 
next-nearest-neighbor bonds. We also have: $H^{00}_{BB}({\bf k})=-H^{00}_{AA}({\bf k}) $ and $H^{\alpha\alpha}_{BB}({\bf k})=-H^{\alpha\alpha}_{AA}({\bf k}) $.
From the mean-field hamiltonian derived in Eq. (\ref{eq:hamflux})
it is evident that the three-spin terms just modify the bare Heisenberg amplitudes of the $\pi$-QSL solution. It is worth 
noting that even a pure $H_\phi$ model with 
infinitesimally small non-zero $J_\phi$ will 
induce a $\pi$-QSL on the square lattice.

\section{Magnetic order from exact diagonalization}

The various magnetic orders in the models considered are investigated
by evaluating the ED static spin structure factor:
\begin{equation}
S({\bf Q})={1 \over N} \sum_{ij} e^{i {\bf Q} \cdot ({\bf R}_i -{\bf R}_j)} \langle {\bf S}_i \cdot {\bf S}_j \rangle,
\end{equation}
where $N$ is the number of sites of the cluster.

\begin{figure}[h!]
	\centering
    \includegraphics[scale=0.42]{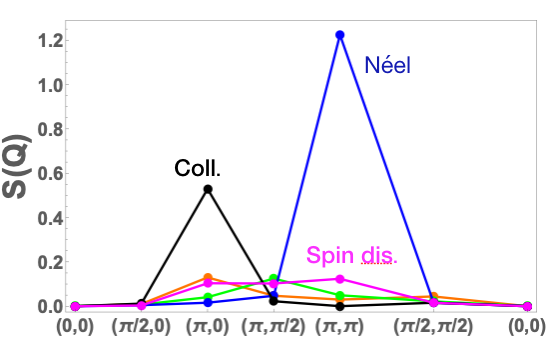}
		\caption{Effect of $J_\phi$ and $D$ on magnetic order of the Heisenberg 
		model on the square lattice. The spin structure factor, $S({\bf Q})$,
		obtained from exact diagonalization on a $4 \times 4$ cluster is 
		shown along the path in the first Brillouin zone 
		of Fig. 1 of the main text. We compare the 
		$S({\bf Q})$ in the Néel, $J_2/J_1=0, J_\phi=D=0$ (blue line) 
		collinear, $J_2/J_1=1, J_\phi=D=0$ (black line), spin disordered, 
		$J_2/J_1=0.6, J_\phi=D=0$ (pink line), 	
		$J_2/J_1=0.6, J_\phi=1,D=0$ (orange line)
		and $D$-state $J_2/J_1=0.6, D=1$ (green line). 
		$S({\bf Q})$ is only evaluated at the points shown so
		the lines are just guides for the eye.
		}
	\label{fig:sq}
\end{figure}

In Fig. \ref{fig:sq} we compare the $S({\bf Q})$ calculated on a $4 \times 4$ cluster  
for different parameters of the $J_1-J_2-D$ and $J_1-J_2-J_\phi$ models.
The Néel and collinear phases display a large clear structure at their
corresponding $(\pi,\pi)$ and $(\pi,0)$ ordering wavevectors, respectively.
This behavior is in contrast to the structureless $S({\bf Q})$ 
of the possible spin disordered phase (around $J_2/J_1=0.6$) or the
weak structure around $(\pi,0)$ ($(\pi,\pi/2)$) of the $J_\phi$ ($D$) 
dominated phase. The fact that these three last phases 
display comparable value of $S({\bf Q})$ indicate that they are disordered
or very weakly magnetically ordered.

\section{Dzyaloshinskii-Moriya contribution}

In this section, we discuss the phase diagram of the $J_1-J_2-D$ model in the presence of a spin-orbit coupling that forces the spin rotations to be coupled with real-space symmetry transformations. The Dzyaloshinskii-Moriya (DM) term is defined as:
\begin{eqnarray}
    H_\text{DM} &=& \sum_{\langle i,j \rangle}
    \vec{D}_{ij} \cdot \left(  \vec{S}_i \cross \vec{S}_j \right)
\end{eqnarray}
where $\vec{D}_{ij}$ are the DM vectors. As said in the paper, we have considered fives compatible DM vector choices, YBCO, LSCO-LTO , LSCO-LTT for the ones compatible with cuprates, and the ones expected to realize rashba-like and dresselhaus-like SO as observed for electronic systems, considered by [\onlinecite{Hotta2019}]. The corresponding $\vec{d}_i$ vectors as displayed  in Fig.~1 of the paper are defined as: (i) YBCO with $\vec{d}_1=\vec{d}_3=(d,0,0)$ and $\vec{d}_2=\vec{d}_4=(0,-d,0)$, (ii) LSCO-LTO with $\vec{d}_1=-\vec{d}_3=(d,0,0)$ and $\vec{d}_2=-\vec{d}_4=(0,-d,0)$, (iii) LSCO-LTT with $\vec{d}_1=-\vec{d}_3=(0,d,0)$ and $\vec{d}_2=-\vec{d}_4=(0,d,0)$, (iv) \textit{Rashba}  with $\vec{d}_1= \vec{d}_3=(0,d,0)$ and $\vec{d}_2= \vec{d}_4=(-d,0,0)$ and (v) \textit{Dresselhaus}  with $\vec{d}_1= \vec{d}_3=(0,-d,0)$ and $\vec{d}_2= \vec{d}_4=(-d,0,0)$.

The effect of the DM term can be considered by employing the MMFT.  Since the DM term only contains two spin operators, using Eq.~\ref{decoupling} it is straightforward to mean field decouple the following terms:
\begin{eqnarray*}
D^x \left( S_i^y S_j^z - S_i^z S_j^y \right), \\
D^y \left( S_i^z S_j^x - S_i^x S_j^z \right), \\ 
D^z \left( S_i^x S_j^y - S_i^y S_j^x \right). \\
\end{eqnarray*}

\begin{figure*}[ht!]
	\centering
    \includegraphics[scale=0.5]{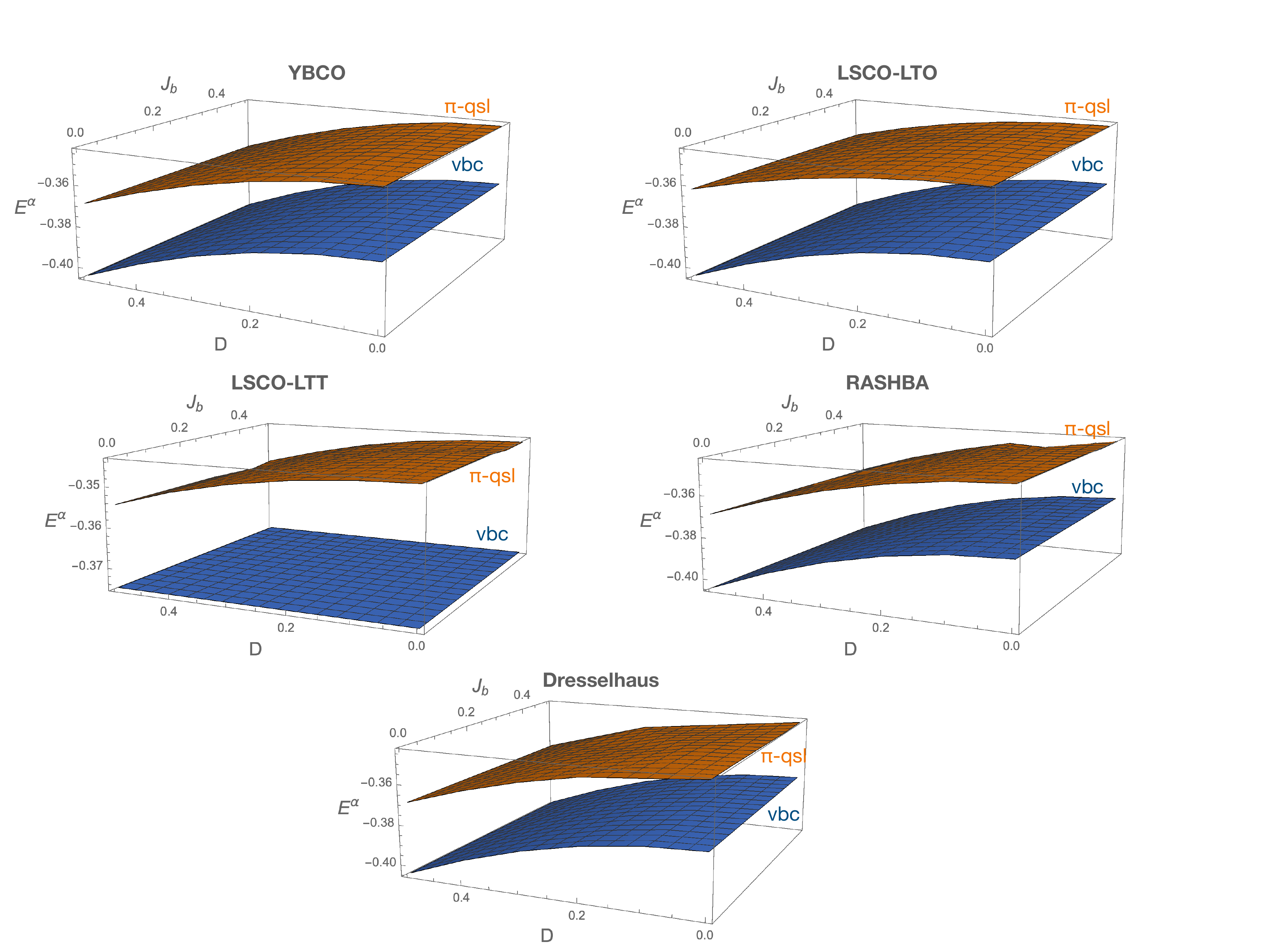}
		\caption{Comparison of the energy of the VBC and the $\pi$-QSL ansatze for the five compatible DM vector choices. The $\pi$-QSL remains always an excited state w.r.t. to VBC one, but the effect of $D$ is to lower its energy hence indicating a positive effect of SO coupling onto this state.}
	\label{fig:compmmftphases}
\end{figure*}

 For each case, we have systematically compared the energies of the 2 most relevant ansatze in the MMFT, namely the VBC state and the $\pi$-QSL state. This is displayed in Fig.~\ref{fig:compmmftphases}.
 
Three important features can be noted. First, thr $\pi$-QSL is always a stable anstaz of our MMFT. Second, the VBC is always lower in energy than the $\pi$-QSL, whatever the choice of the DM vectors. Finally, the $\pi$-QSL is also always lowered w.r.t. $D$ indicating that the effect of SO coupling is positive, in the sense that it continues to favor it even though it is an excited state.

To further understand the role of the DM term, we have solved the $J_1-J_2-D$ Hamiltonian by using exact diagonalizations on a 16 site cluster and the Majorana mean field theory for the case compatible with YBCO compounds.
In Fig.\ref{fig:pdgap}, we show a ternary plot with the constraint $J_1+J_2+D=1$, of the first gap obtained by ED. We immediately see five distinct domains separated by zero gap lines (white regions).   \begin{figure*}[t!]
	\centering
    \includegraphics[scale=0.45]{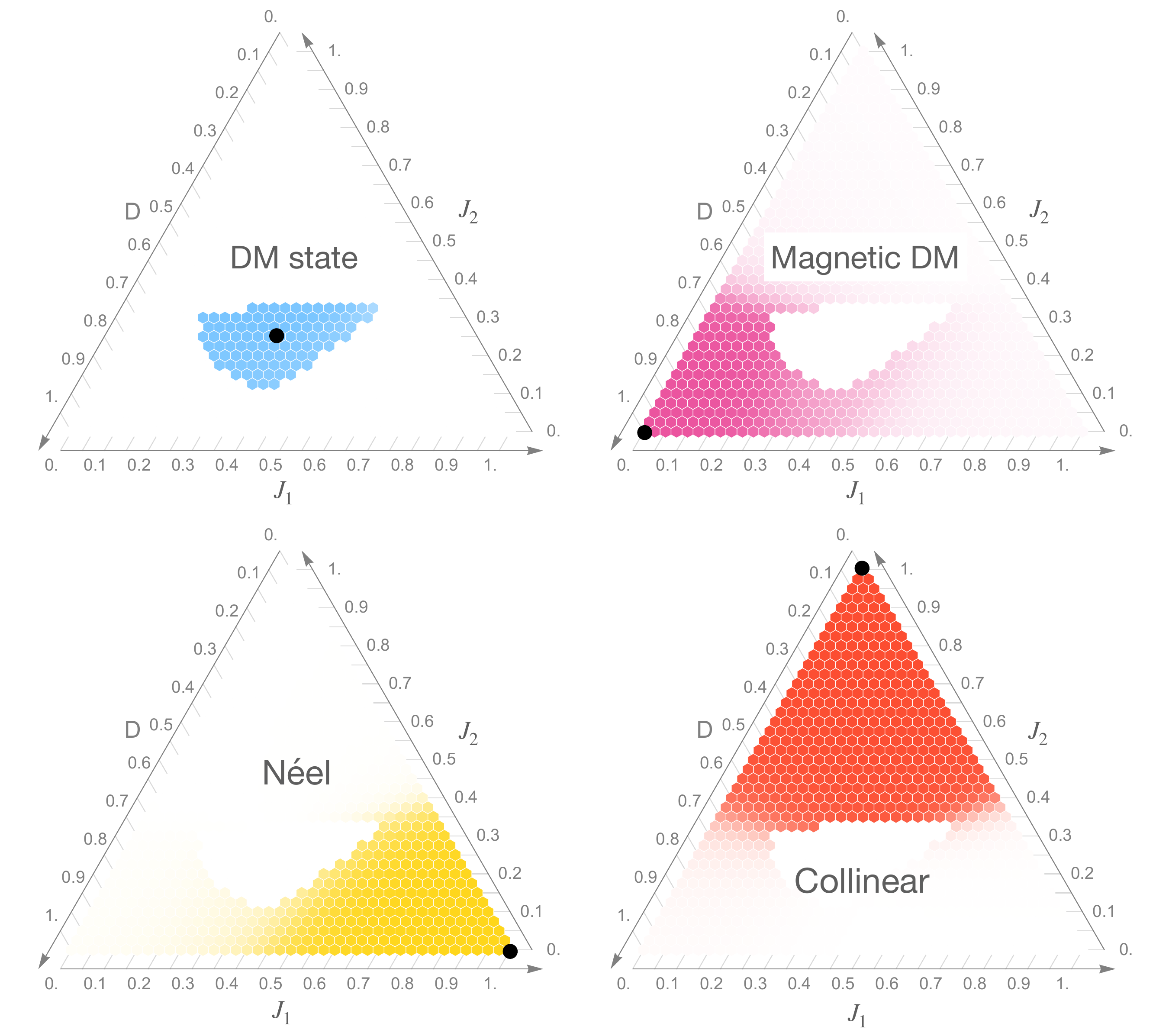}
		\caption{Quantum fidelity computed by ED for 4 different reference sites (black points). A quantum fidelity of 1 corresponds to full colored point while a zero one corresponds to white hexagons. Four clear regions can be clearly identified, as mentioned in Fig.\ref{fig:pdgap}. The disordered phase of the raw $J_1-J_2$ model cannot be clearly identified in such a small cluster, but clearly appears in the gap of Fig.~\ref{fig:pdgap} (extended white region). }
	\label{fig:fidelity}
\end{figure*}
 \begin{figure}[ht!]
	\centering
    \includegraphics[scale=0.4]{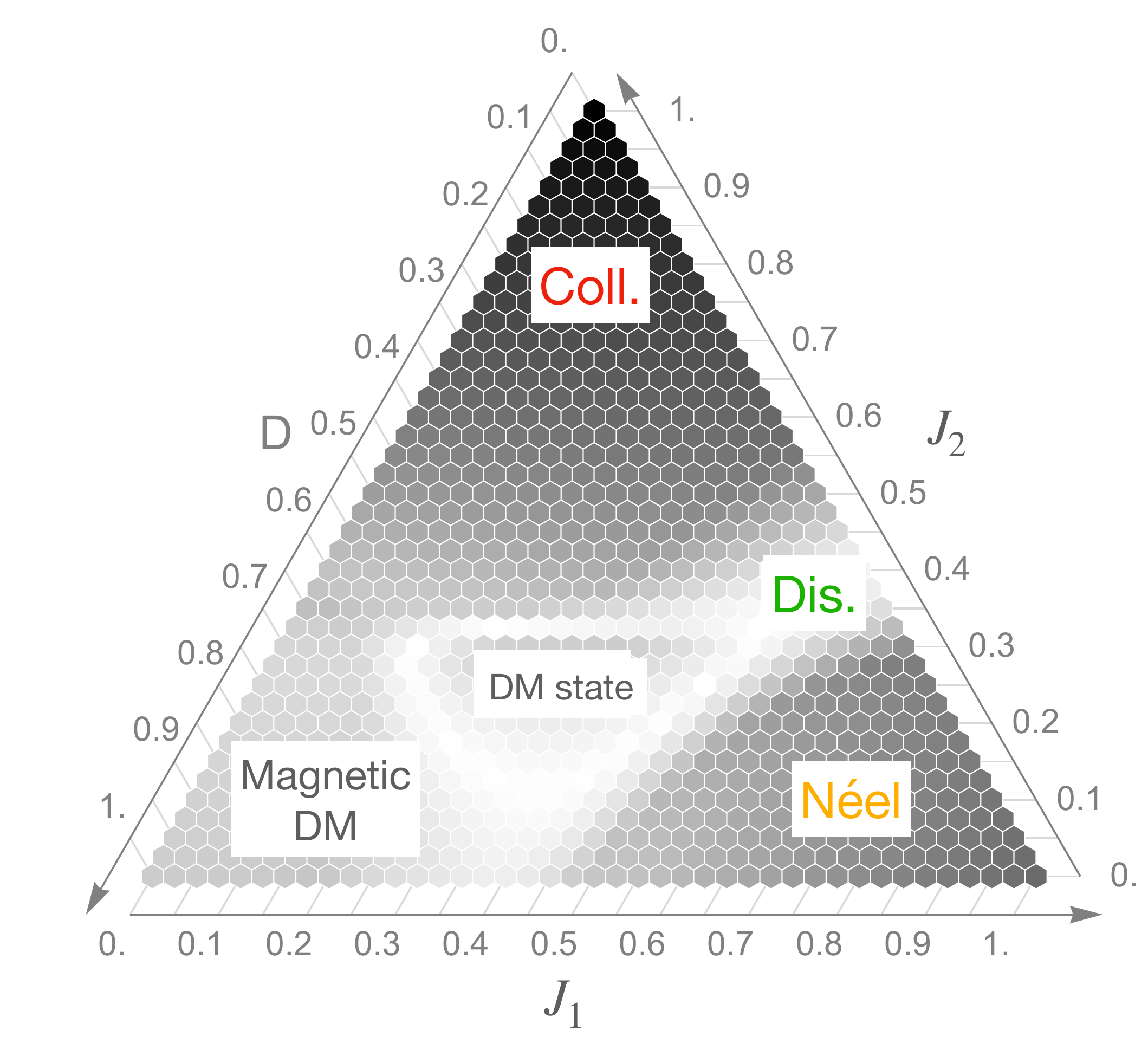}
		\caption{Density map of the gap $\Delta E$ on a ternary plot satisfying $J_1 + J_2 + D = 1$ obtained by ED on the 16-site cluster. The more white the density, the smaller the gap. Five phases are observed, the Néel, the disordered and the collinear phases from the original $J_1-J_2$ model that largely extend in the phase diagram, an ordered phase driven by $D$ called magnetic DM and a trivial and possibly disordered intermediate phase dubbed DM state, delimited by frontiers of zero gap (white lines).}
	\label{fig:pdgap}
 \end{figure}  
Along the $D=0$ line, we recover the Néel region (close to $J_1=1$), the intermediate disordered phase ($ 0.4 <J_2 < 0.6 $) and the collinear phase at higher $J_2$. In order to characterize these phases we have calculated the quantum fidelity $| \langle \psi_\text{ref} | \psi_0 \rangle |$ in the whole parameter space by starting from 4 reference states $| \psi_\text{ref} \rangle $ chosen deep in their corresponding expected parameter regions. They are indicated by black points in Fig.\ref{fig:fidelity} for each of the 4 panels, and the value of the quantum fidelity, ranging from 0 to 1, scales from white (0) to full colored (1) hexagons respectively. 

Strikingly, already on such a limited cluster, the fidelity gives strong indications about 
the phase boundary locations. For instance, from the overlap with the pure Néel state obtained at $J_1 = 1$, we find that the Néel region extends in the lower right part of the ternary plot (yellow points). A very small, yet finite, fidelity can be found in the DM region as $J_1$ decreases, but considering that the Néel region is unambiguously characterized  
by a fidelity close to one then, along the $J_2 = 0$ line, the Néel region ends
around $J_1 \simeq 0.4$. What is remarkable here is that the Néel, the DM and the collinear regions are well separated even when the transition is second order (no level crossings).\\ As previously mentioned, starting from a different reference state allows us to span 
entire regions of this model with close-to-one fidelities. This is in particular the case for the intermediate region (blue centered domain of the upper-left panel of Fig.~\ref{fig:fidelity}) delimited by a frontier of zero energy gap as can be observed in Fig.\ref{fig:pdgap}. This region (the blue one in Fig.\ref{fig:fidelity}) is gapped and
possibly magnetically disordered since the spin structure factor, 
$S({\bf Q})$, is rather featureless displaying 
only weak structures as shown in Fig. \ref{fig:sq}. We have denoted this intermediate 
phase the DM state in the main text. Finally, at large $D$, the system enters a fourth 
phase which is magnetically ordered. This can be interpreted as a 
quantum version of the magnetic DM state obtained with the MMFT approach. 
To the best of our knowledge, such intermediate possibly disordered 
DM state has never been reported so far.

\section{Details on the temperature dependence of thermal Hall conductivity }

\begin{figure}[ht!]
	\centering
	\includegraphics[width=8cm,clip]{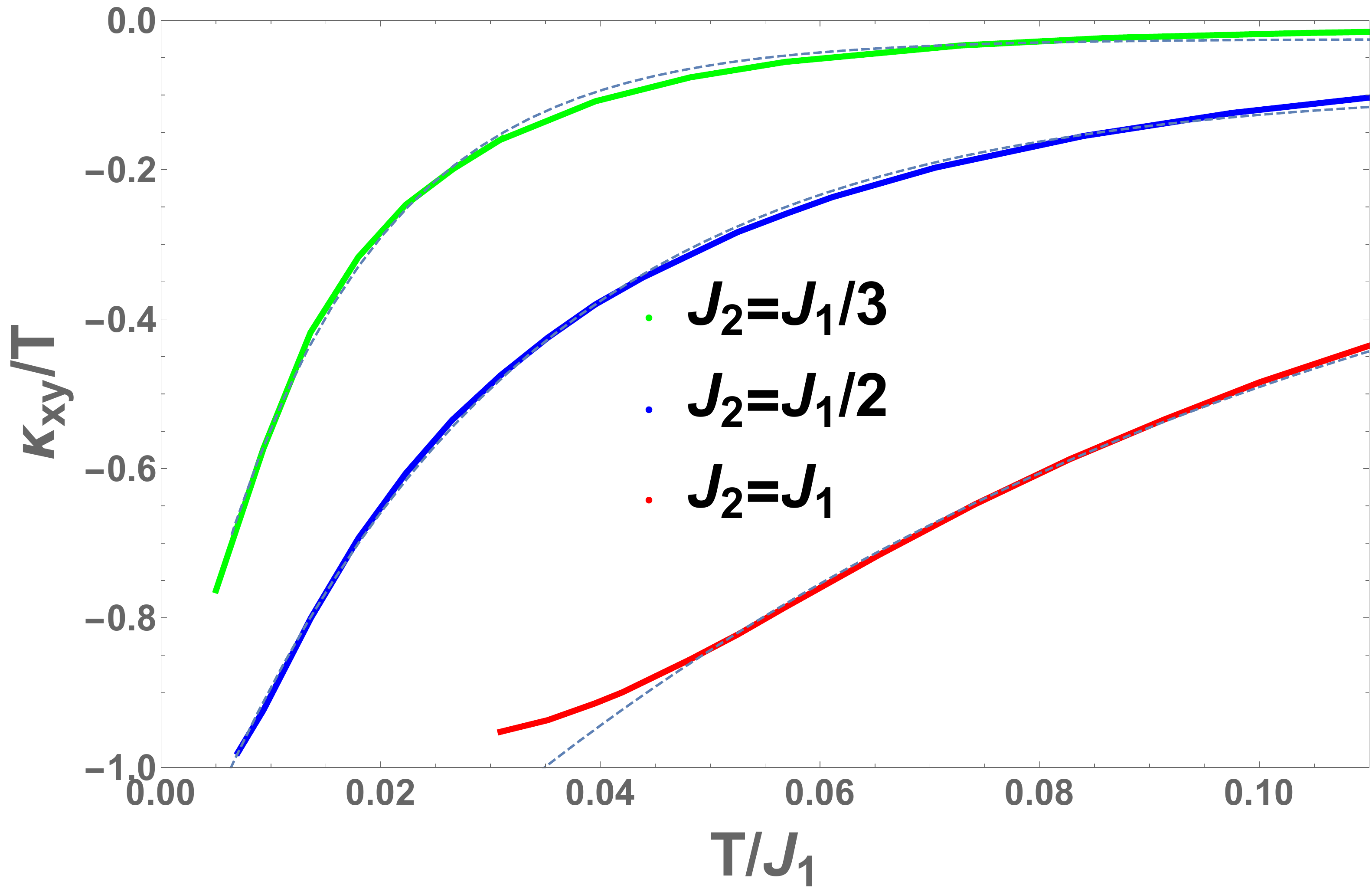}
	\caption{Exponential temperature dependence of the thermal Hall conductivity. Fits of our calculated MMFT thermal Hall conductivity to: $\kappa_{xy}/T \sim A+ B e^{-T/T_0}$ are 
	shown. } 
	\label{fig:fits}
\end{figure}

In the main text we have stated that the thermal Hall conductivity can be fitted in a broad temperature range to a exponential form $\kappa_{xy}/T \propto e^{-T/T_0}$. Such exponentical forms are observed experimentally in an intermediate temperature range \cite{Taillefer2020a}.
Here we provide details about these fits. Our theoretical calculations can be nicely fitted to the following expressions: 
$\kappa_{xy}/T \sim A+ B e^{-T/T_0}$  as shown in Fig. \ref{fig:fits}. For $J_2/J_1=1/3$ the fit is: $\kappa_{xy}/T \sim -0.02508 -1.02225 e^{-T/0.015}$,  for $J_2/J_1=0.5$ the fit is:  $\kappa_{xy}/T \sim -0.0905446-1.13263 e^{-T/0.029}$ and for $J_2/J_1=1$ the fit is: $\kappa_{xy}/T \sim -0.0821182 -1.41715 e^{-T/0.08045}$. Using appropriate parameters 
for the cuprates $J_1 \sim 0.1 eV$ would lead to the temperature scales: $T_0(J_2/J_1=1/3) \sim 18 $K,
$T_0(J_2/J_1=1/2) \sim 35 K$ and $T_0(J_2/J_1) \sim 93.4$ K. 

We can compare our theoretical MMFT results with the 2D thermal conductivity obtained from experiments. Experimental data of La$_2$CuO$_4$ and Sr$_2$CuO$_2$Cl$_2$ 
has been found to have the $T$-dependence : $\kappa^{expt}_{xy}/T \sim (-0.0132  - 3.234 e^{-T/T^{expt}_0}) \times 10^{-9}
{mW \over K^2}$ with $T^{expt}_0\sim 17 K$. The theoretical curve with closest $T_0$ corresponds (taking $J_1 \sim 0.1$ eV) to $J_2/J_1=1/3$ ($T_0 \sim 18$ K) and has the $T$-dependence: $\kappa_{xy}/T \sim (-0.01618 - 1.2073 e^{-T/T_0} ) \times 10^{-9} {mW \over K^2}$. 
Since the cuprates are typically in the 
low $J_2/J_1$ parameter regime, we conclude that our 
MMFT predictions for $\kappa_{xy}$ are consistent with
available experimental data. 

